\documentclass[twocolumn,PRB]{revtex4}
\usepackage{type1cm}
\usepackage[dvipdfm]{graphicx}

\usepackage{amsmath,amssymb,bm,amsthm}

\def\beq{\begin{equation}}
\def\eeq{\end{equation}}
\def\nbeq{\begin{equation*}}
\def\neeq{\end{equation*}}

\def\<{\langle}
\def\>{\rangle}
\def\rt#1{\sqrt{\mathstrut #1}}
\def\Tr{{\rm Tr}}
\def\tr{{\rm tr}}
\renewcommand{\d}{\partial}
\newcommand{\arginf}{\operatornamewithlimits{arginf}}
\newcommand{\argsup}{\operatornamewithlimits{argsup}}

\begin{document}
\title{Equilibrium Properties of Quantum Spin Systems with Non-additive Long-Range Interactions}
\author{Takashi Mori}
\affiliation{
Department of Physics, Graduate School of Science,
The University of Tokyo, Bunkyo-ku, Tokyo 113-0033, Japan
}
\begin{abstract}
We study equilibrium states of quantum spin systems with nonadditive long-range interactions 
by adopting an appropriate scaling of the interaction strength, {\it i.e.}, the so called Kac prescription.
In classical spin systems, it is known that the equilibrium free energy is obtained by minimizing the free energy functional over the coarse-grained magnetization.
Here we show that it is also true for quantum spin systems.
From this observation, it is found that
when the canonical ensemble and the microcanonical ensemble are not equivalent in some parameter region,
it is not necessarily justified to replace the actual long-range interaction by the infinite-range interaction (Curie-Weiss type interaction).
On the other hand, in the parameter region where the two ensembles are equivalent,
this replacement is always justified.
We examine the Heisenberg XXZ model as an illustrative example
and discuss the relation to experiments.
\end{abstract}
\maketitle
\section{Introduction}

Long-range interacting systems have attracted much attention because of a wide variety of physical systems 
and peculiarities of its dynamics and thermodynamics~\cite{Campa_review,Les_Houches_long}.
Recent experimental achievement allows us to access artificial quantum lattice systems where the coupling parameters can be controlled.
Dipolar gases in optical traps are suggested to be possible experimental realization of long-range interacting lattice systems.
Moreover, it has been suggested~\cite{ODell2000} that we can realize artificial gravitational systems in laboratory
by irradiating off-resonant laser beam onto atoms in a Bose-Einstein condensate.
Theoretically, long-range interactions exhibit rich equilibrium and non-equilibrium behavior 
like the ensemble inequivalence, negative specific heat, and the absence of thermalization.
These properties are atypical for short-range interacting systems.
Therefore, studying long-range interacting systems will lead us to deepen our fundamental understanding of statistical mechanics.

Those peculiar features of long-range interacting systems have been understood by using mean-field models~\cite{Campa_review}.
In mean-field models, the interaction between particles or spins is assumed to be independent of distance.
Since exact results are available in mean-field models, they are important theoretically although they appear to be too simplified.
Thus exploring the condition on the realization of predictions by mean-field models should be actively proceeded.

Kastner~\cite{Kastner2010,Kastner_statmech2010} examined quantum spin systems with an infinite-range ferromagnetic interaction.
He demonstrated the ensemble inequivalence of these systems.
In this paper, more general long-range interactions are treated;
ferromagnetic interactions whose range is comparable with the system size.
An important class of these interactions is the power-law potential $\phi(r)\sim -1/r^{\alpha}$, where $r$ is the spatial distance between spins
and $\alpha$ satisfies $0\leq\alpha<d$ ($d$ is the spatial dimension).
It has been believed from the analysis of exact partition functions or the numerical calculations of some models 
that thermodynamic behavior does not depend on $\alpha$ as long as it is less than $d$~\cite{Cannas2000,Tamarit2000,Campa2003,Barre2005}.
As $\alpha=0$ corresponds to the mean-field model, 
it implies that the results of the mean-field model might be exact for general long-range interacting systems 
as long as $\phi(r)\sim 1/r^{\alpha}$, and $0\leq\alpha<d$.

In the previous papers~\cite{Mori2010_analysis,Mori2011_instability,Mori2012_microcanonical}, 
it was shown that for classical spin systems, the free energy does not depend on $\alpha$ in the canonical ensemble with non-conserved order parameters,
but $\alpha$ becomes relevant in some parameter region when there is a constraint of the value of the energy or the magnetization.
A sufficient and a necessary condition for $\alpha$ to be relevant are derived in the previous works.
In this paper, we show that it is also true in quantum spin systems.

An immediate conclusion is that
if the inequivalence of the canonical and the microcanonical ensembles predicted by the analysis of the mean-field model
can be observed in experiments,
the violation of exactness of the mean-field theory will be also observed in the microcanonical ensemble.
Inhomogeneous spin configurations in equilibrium are its sign.
Although it is known that the mean-field models reproduce many properties of general long-range interacting systems even quantitatively,
the infinite-range (or Curie-Weiss type) interaction does not become an adequate idealization of actual long-range interactions in such a situation.

This paper is organized as follows.
In Sec.~\ref{sec:model}, we introduce the model.
In Sec.~\ref{sec:cg}, an important notion of coarse-graining is explained.
In Sec.~\ref{sec:classical}, we express the free energy only by classical variables.
In Sec.~\ref{sec:variational}, the variational expression of the free energy is derived.
In Sec.~\ref{sec:exactness}, we mention some results deduced from the previous sections,
namely that the free energy is independent of the precise form of the interaction potential 
as long as there is no constraint such as fixed value of the energy or the magnetization. 
On the other hand, the free energy under the constraint of the fixed value of the magnetization and the microcanonical entropy
may depend on the interaction potential.
This result is the same as in classical spin systems with long-range interactions.
In Sec.~\ref{sec:ex}, we investigate the spin-1/2 XXZ model in the canonical and the microcanonical ensemble as an illustrative example.
In Sec.~\ref{sec:summary}, we summarize our results.

\section{Model}
\label{sec:model}

We study systems expressed by the following Hamiltonian on $d$-dimensional regular lattice:
\beq
H=-\frac{1}{2s^2}\sum_{ij}^N\gamma^d\phi(\gamma \bm{r}_{ij})\sum_{a=x,y,z}\lambda_as_i^as_j^a-\frac{\vec{h}}{s}\cdot\sum_i^N\vec{s}_i.
\label{eq:H}
\eeq
We focus on the ferromagnetic interaction, $\lambda_a\geq 0$ and $\phi(\cdot)\geq 0$.
The parameter $\gamma$ corresponds to the inverse of the interaction range and 
we take the limit of $\gamma\rightarrow 0$ appropriately (see below).
$\phi(\cdot)$ is the interaction potential and we will give the condition of $\phi$ later.
$\vec{s}_i$ is the spin operator at site $i$ whose length $s$ is independent of particular site $i$ and $s\in\{1/2,1,3/2,\dots\}$.
Normalizations $1/s^2$ for the interaction strength and $1/s$ for the magnetic field $\vec{h}$ are not essential.
With these normalizations, the system is reduced to a classical continuous spin system in the limit of $s\rightarrow\infty$.
The lattice interval is set to be unity and $L$ denotes the length of the system (the number of spins is $L^d=N$).
The position of site $i$ is denoted by $\bm{r}_i\in \mathbb{Z}^d\cap [1,L]^d$.
The distance between two sites $i$ and $j$ is denoted by $\bm{r}_{ij}$.

We take the two limits; $L\rightarrow\infty$ and $\gamma\rightarrow 0$.
In the present paper, we consider either of the following limiting procedures:
\begin{itemize}
\item[(i)] {\bf nonadditive limit}: 
the condition for $\phi(\bm{r})$ is that $\phi(\bm{r})\geq 0$ and $\int_{\bm{r}\in[0,\gamma L]^d}\phi(\bm{r})d^dr=1$.
We take the limit $\gamma\rightarrow 0$, $L\rightarrow\infty$ with $\gamma L$ fixed.
Note that it includes power-law potentials $\phi(\bm{r})\sim 1/r^{\alpha}$, $0\leq\alpha<d$.
\item[(ii)] {\bf van der Waals limit}:
the potential $\phi(\bm{r})$ satisfies $\phi(\bm{r})\geq 0$ and $\int\phi(\bm{r})d^dr<+\infty$. 
We take the van der Waals limit, that is $\gamma\rightarrow 0$ after $L\rightarrow\infty$.
\end{itemize}
The case (i) corresponds to nonadditive interactions,
namely the interaction range is comparable with the system size.
On the other hand, in case (ii), the interaction range is much longer than the lattice interval but much shorter than the system size $L$.
Hence, two limiting procedures treat different situations.

Lebowitz and Penrose~\cite{Lebowitz-Penrose1966} proved that in the van der Waals limit,
the equation of state of the system is equal to the mean-field equation of state 
(van der Waals like equation) with the so called Maxwell's equal area rule (or referred to be as the Maxwell construction).
In the case of (ii), therefore, the free energy is independent of the precise form of the interaction potential $\phi(\cdot)$.
Extension to quantum many particle systems was done by Lieb~\cite{Lieb1966}.
The perturbative analysis around $\gamma=0$ was suggested 
to understand the properties of a system with long but finite range interaction ($\gamma>0$)~\cite{Brout1960}.
Even in the critical region, some exact results were obtained along this line~\cite{Bricmont1982}.

On the other hand, the case of (i) covers much {\it longer} interactions than the van der Waals limit. 
Similarly to the case of (ii), it has been indicated that the free energy does not depend on the interaction potential, 
and it is equivalent to the mean-field free energy {\it without} the Maxwell construction in the nonadditive limit.
It is called the exactness of the mean-field theory~\cite{Cannas2000}.
Therefore, we can regard the exactness of the mean-field theory in the nonadditive limit 
is a generalization of the Lebowitz-Penrose theorem in the van der Waals limit.
Actually, it is correct in classical spin systems 
unless there are some constraints such as fixed energy or fixed magnetization~\cite{Mori2010_analysis,Mori2011_instability,Mori2012_microcanonical},
as mentioned in the Introduction.

\section{Coarse graining}
\label{sec:cg}

In this section, we introduce a kind of coarse graining and show that the equilibrium free energy is exactly given by the free energy of the coarse-grained model.
We divide the system into many cells which are large enough to contain many spins, but small compared to the interaction range.
The length of each cell is denoted by $l$ and cells are labeled by ${\rm C}_p$, $p=1,2,\dots,(L/l)^d$.
The resultant spin of a cell ${\rm C}_p$ is denoted by $\vec{S}_p\equiv\sum_{i\in {\rm C}_p}\vec{s}_i$.
Eigenvalues of $(\vec{S}_p)^2$ and $S_p^z$ are denoted by $J_p(J_p+1)$ and $M_p$, respectively (we set $\hbar$ to be unity).
In each cell ${\rm C}_p$, there are $l^d$ spins, so they are properly labeled by $\vec{s}_{p(1)},\vec{s}_{p(2)},\dots,\vec{s}_{p(l^d)}$.
We define $n$-partial resultant spin as
\beq
\vec{S}_p^{(n)}\equiv\sum_{i=1}^n\vec{s}_{p(i)},
\eeq
and eigenvalues of $(\vec{S}_p^{(n)})^2$ are denoted by $J_p^{(n)}(J_p^{(n)}+1)$.
Then the full Hilbert space is spanned by the tensor product of eigenstates 
$|J_p,M_p,\Gamma_p^{(l^d)}\>$ of each cell, $\bigotimes_{p=1}^{(L/l)^d}|J_p,M_p,\Gamma_p^{(l^d)}\>$.
Here $$\Gamma_p^{(l^d)}\equiv\{0,J_p^{(1)},J_p^{(2)},\dots,J_p^{(l^d)}=J_p\}$$ is the ``trajectory'' of $J_p^{(n)}$, $n=1,2,\dots,l^d$.
It is explained later in more detail.

When the interaction is long-range, it is expected that the Hamiltonian can be expressed only by the resultant spin operators $\{ \vec{S}_p\}$ approximately.
Moreover, typically $J_p$ is considered to be very large when $l\gg 1$,
so it is naturally expected that all the $\vec{S}_p$, $p=1,2,\dots,(L/l)^d$, can be regarded as classical spins whose length is $J_p$.

We shall construct the above argument rigorously.
The coarse-grained Hamiltonian is introduced by
\beq
\tilde{H}=-\frac{1}{2s^2}\sum_{p,q}^{(L/l)^d}\gamma^d\phi_{pq}\sum_{a=x,y,z}S_p^aS_q^a-\frac{\vec{h}}{s}\cdot\sum_p^{(L/l)^d}\vec{S}_p,
\label{eq:cg_H}
\eeq
where
\beq
\phi_{pq}\equiv\frac{1}{l^{2d}}\sum_{i\in {\rm C}_p}\sum_{j\in {\rm C}_q}\phi(\gamma\bm{r}_{ij}).
\eeq
The above coarse-grained Hamiltonian is expressed only by the resultant spin operators of each cell.
The original Hamiltonian is well approximated by this coarse-grained one.
In order to see it, consider the operator
\beq
g\equiv\frac{1}{L^d}(H-\tilde{H}).
\eeq
We define the norm of the operator $A$ by
\beq
\| A\|\equiv \sup_{\psi, \<\psi|\psi\>=1}\left|\<\psi|A|\psi\>\right|.
\eeq
The norm of $g$ then is given by
\begin{align}
\| g\| =&\frac{L^d}{2\gamma^ds^2}\left\|\sum_{p,q}^{(L/l)^d}\sum_{i\in {\rm C}_p}\sum_{j\in {\rm C}_q}(\phi(\gamma\bm{r}_{ij})-\phi_{pq})
\sum_{a=x,y,z}\lambda_as_i^as_j^a\right\|
\nonumber \\
\leq&\frac{L^d}{2\gamma^d}\sum_{p,q}^{(L/l)^d}\sum_{i\in {\rm C}_p}\sum_{j\in {\rm C}_q}|\phi(\gamma\bm{r}_{ij})-\phi_{pq}|
\nonumber \\
&\qquad\times\frac{\left\|\sum_{a=x,y,z}\lambda_as_i^as_j^a\right\|}{s^2}.
\end{align}
Here $\left\|\sum_a\lambda_as_i^as_j^a\right\|/s^2\leq\lambda_{\rm max}\equiv\max_{a=x,y,z}\lambda_a$, therefore
\beq
\| g\|\leq\frac{(\gamma L)^d\lambda_{\rm max}}{2}
\sum_{p,q}^{(L/l)^d}\sum_{i\in {\rm C}_p}\sum_{j\in {\rm C}_q}|\phi(\gamma\bm{r}_{ij})-\phi_{pq}|\equiv g_{\rm cl}.
\eeq
In Ref.~\cite{Mori2011_instability}, it is shown that $$\lim_{l\rightarrow\infty}\lim_{L\rightarrow\infty}g_{\rm cl}=0$$ for the nonadditive limit
and $$\lim_{l\rightarrow\infty}\lim_{\gamma\rightarrow 0}\lim_{L\rightarrow\infty}g_{\rm cl}=0$$ for the van der Waals limit.
Therefore, the original Hamiltonian is well approximated by the coarse-grained one.
For convenience, we use the same notation ``${\rm Lim}$'' for the nonadditive limit and the van der Waals limit.
The above property then implies
\beq
{\rm Lim}\| g\|=0.
\label{eq:lim_g}
\eeq

By using this property, we shall show that the free energy per spin is exactly the same as that given by the coarse-grained Hamiltonian.
The free energy is defined by
\beq
f(\beta,\vec{h})=-\frac{1}{L^d\beta}\ln\Tr e^{-\beta H}.
\eeq
Here $\beta=1/T$ is the inverse temperature.
The coarse-grained free energy is given by
\beq
\tilde{f}(\beta,\vec{h})=-\frac{1}{L^d\beta}\ln\Tr e^{-\beta\tilde{H}}.
\eeq

From now on, we show that ${\rm Lim}|f-\tilde{f}|=0$.
We use the Bogoliubov-Peierls inequality,
\beq
\Tr e^{-\beta H}\geq \sum_ie^{-\beta \< i|H|i\>},
\label{eq:Peierls}
\eeq
where $\{|i\>\}$ is arbitrary orthonormal set.
From $H=\tilde{H}+L^dg$ and Eq.~(\ref{eq:Peierls}), we obtain
\beq
\Tr e^{-\beta H}\geq \sum_ie^{-\beta \< i|\tilde{H}|i\>}e^{-\beta L^d\< i|g|i\>}.
\label{eq:BP_bound}
\eeq
By choosing the eigenstates of $\tilde{H}$ as $\{|i\>\}$, we have
\beq
\Tr e^{-\beta H}\geq \sum_ie^{-\beta \tilde{E}_i}e^{-\beta L^d\< i|g|i\>}\geq e^{-\beta L^d\| g\|}\Tr e^{-\beta\tilde{H}},
\eeq
where $\{\tilde{E}_i\}$ are eigenvalues of $\tilde{H}$.

On the other hand, the upper bound of the partition function $\Tr e^{-\beta H}$ is obtained 
by using the Golden-Thompson inequality~\cite{Golden1965,Thompson1965},
\beq
\Tr e^{-\beta H}=\Tr e^{-\beta\tilde{H}-\beta L^dg}\leq \Tr e^{-\beta\tilde{H}}e^{-\beta L^dg}.
\eeq
We have
\begin{align}
\Tr e^{-\beta\tilde{H}}e^{-\beta L^dg}&=\sum_ie^{-\beta\tilde{E}_i}\< i|e^{-\beta L^dg}|i\>
\nonumber \\
&\leq \sum_ie^{-\beta\tilde{E}_i}\| e^{-\beta L^dg}\|
\nonumber \\
&=e^{\beta L^d\| g\|}\Tr e^{-\beta\tilde{H}}.
\label{eq:GT_bound}
\end{align}
From Eq.~(\ref{eq:BP_bound}) and Eq.~(\ref{eq:GT_bound}), we obtain
\beq
e^{-\beta L^d\| g\|}\leq \frac{\Tr e^{-\beta H}}{\Tr e^{-\beta\tilde{H}}}\leq e^{\beta L^d\| g\|},
\eeq
and thus
\beq
|f-\tilde{f}|\leq \| g\|.
\eeq
From Eq.~(\ref{eq:lim_g}), ${\rm Lim}|f-\tilde{f}|=0$.
Hence we can safely replace $H$ by $\tilde{H}$ to compute the free energy.

\section{Classical bounds}
\label{sec:classical}

The coarse-grained Hamiltonian depends only on the resultant spins $\{\vec{S}_p\}$.
Since there are $l^d$ spins inside a cell and finally we take the limit of $l\rightarrow\infty$,
it is expected that the typical length of resultant spins is huge.
We are then able to replace the quantum mechanical operator $\vec{S}_p$ by the classical vector.
In this section, we develop a theory based on the above argument rigorously.

We consider the Hilbert space of a cell ${\rm C}_p$.
The spins in this cell is labeled by $\vec{s}_{p(i)}$, $i=1,2,\dots,l^d$.
As is mentioned in Sec.~\ref{sec:cg}, the Hilbert space of the cell ${\rm C}_p$ is spanned by the complete orthonormal basis
$\{|J_p,M_p,\Gamma_p^{(l^d)}\>\}$.

We notice that when each spin $\vec{s}_i$ is a spin-$s$ operator,
\begin{align}
J_p^{(k)}\in\left\{ J_p^{(k-1)}+s, J_p^{(k-1)}+s-1, J_p^{(k-1)}+s-2\right.,
\nonumber \\
\left.\dots, J_p^{(k-1)}-s\right\}\bigcap [0,\infty).
\label{eq:process}
\end{align}
The ``trajectory'' with length $l^d$ is defined by $$\Gamma_p^{(l^d)}(0\rightarrow J_p^{(l^d)})\equiv\left\{ 0,J_p^{(1)},\dots,J_p^{(l^d)}\right\},$$
where $\{J_p^{(k)}\}$ satisfies Eq.~(\ref{eq:process}).
We will omit $(0\rightarrow J_p)$ when there is no confusion.

The set of trajectories of all $\Gamma_p^{(l^d)}(0\rightarrow J_p)$ with a fixed $J_p$ is denoted by ${\cal T}_{l^d}(J_p)$.
Moreover, we define the {\it weight function} $W(J_p)$ by
\beq
W(J_p)\equiv \left[\text{the number of elements of ${\cal T}_{l^d}(J_p)$}\right].
\eeq
Explicit calculation of the weight function is given in Appendix~\ref{sec:W}.

The identity operator in the Hilbert space of a cell ${\rm C}_p$ is expressed by
\beq
1_p=\sum_{J_p}\sum_{M_p=-J_p}^{J_p}\sum_{\Gamma_p^{(l^d)}\in {\cal T}_{l^d}(J_p)}|J_p,M_p,\Gamma_p^{(l^d)}\>\< J_p,M_p,\Gamma_p^{(l^d)}|.
\eeq
Because the coarse-grained Hamiltonian depends only on the redundant spins $\vec{S}_p$,
the partition function of the coarse-grained Hamiltonian is written as
\begin{align}
\Tr e^{-\beta\tilde{H}}&=\prod_{p=1}^{(L/l)^d}\left\{\sum_{J_p}\sum_{M_p=-J_p}^{J_p}\sum_{\Gamma_p^{(l^d)}\in{\cal T}_{l^d}(J_p)}\right\}
\nonumber \\
\times&\left[\bigotimes_{p=1}^{(L/l)^d}\< J_p,M_p,\Gamma_p^{(l^d)}|\right]
e^{-\beta\tilde{H}}\left[\bigotimes_{p=1}^{(L/l)^d}|J_p,M_p,\Gamma_p^{(l^d)}\>\right]
\nonumber \\
&=\prod_{p=1}^{(L/l)^d}\left\{\sum_{J_p}\sum_{M_p=-J_p}^{J_p}W(J_p)\right\}
\nonumber \\
\times&\left[\bigotimes_{p=1}^{(L/l)^d}\< J_p,M_p|\right]e^{-\beta\tilde{H}}\left[\bigotimes_{p=1}^{(L/l)^d}|J_p,M_p\>\right] \nonumber \\
&\equiv \prod_{p=1}^{(L/l)^d}\left\{\sum_{J_p}W(J_p)\right\}\Tr_{\{J_p\}}e^{-\beta\tilde{H}}.
\label{eq:partition}
\end{align}
Here $|J_p,M_p\>$ is an arbitrary state vector such that 
$$|J_p,M_p\>\in\{|J_p,M_p,\Gamma_p^{(l^d)}\>, \Gamma_p^{(l^d)}\in{\cal T}_{l^d}(J_p)\}.$$
For instance, we choose $|J_p,M_p\>$ as
\beq
|J_p,M_p\>=\frac{1}{\sqrt{W(J_p)}}\sum_{\Gamma_p^{(l^d)}\in{\cal T}_{l^d}(J_p)}|J_p,M_p,\Gamma_p^{(l^d)}\>.
\eeq
The symbol $\Tr_{\{J_p\}}\equiv\tr_{J_1}\tr_{J_2}\dots\tr_{J_{(L/l)^d}}$ means the trace over the subspace of fixed values of $\{J_p\}$,
\beq
\tr_{J_p}A\equiv\sum_{M_p=-J_p}^{J_p}\<J_p,M_p|A|J_p,M_p\>.
\eeq



Lieb~\cite{Lieb1973} proved that for the general Hamiltonian such as Eq.~(\ref{eq:cg_H}), if $\phi_{pp}\geq 0$,
\beq
Z^{\rm cl}(\{J_p\})\leq\frac{\Tr_{\{J_p\}}e^{-\beta\tilde{H}}}{\prod_p(2J_p+1)}\leq Z^{\rm cl}(\{J_p+1\}),
\label{eq:Lieb1973}
\eeq
where $Z^{\rm cl}(\{J_p\})$ is the {\it classical} partition function with spins $\{J_p\}$, namely,
\begin{align}
Z^{\rm cl}(\{J_p\})\equiv&\int \frac{d\Omega_1}{4\pi}\int \frac{d\Omega_2}{4\pi}\dots\int \frac{d\Omega_{(L/l)^d}}{4\pi}
\nonumber \\
&\times\exp\left[-\beta\tilde{H}^{\rm cl}(\{J_p,\theta_p,\phi_p\})\right].
\end{align}
Here, the classical Hamiltonian is defined by the same form of Eq.~(\ref{eq:cg_H}) but the spin operators $\vec{S}_p$ are replaced by
the classical vectors, $$\vec{S}_p\rightarrow J_p(\sin\theta_p\cos\phi_p,\sin\theta_p\sin\phi_p,\cos\theta_p).$$
Lieb proved the inequality~(\ref{eq:Lieb1973}) by using a spin coherent state representation.

From Eqs.~(\ref{eq:partition}) and (\ref{eq:Lieb1973}), we obtain the classical bounds
\beq
Z_-\leq Z\leq Z_+,
\eeq
where
\begin{align*}
Z_-\equiv&\sum_{\{J_p\}}{\cal W}(\{J_p\})\int D\Omega\left[\prod_p(2J_p+1)\right]
\nonumber \\
&\times\exp\left[-\beta\tilde{H}^{\rm cl}(\{J_p,\theta_p,\phi_p\})\right], \\
Z_+\equiv&\sum_{\{J_p\}}{\cal W}(\{J_p\})\int D\Omega\left[\prod_p(2J_p+1)\right]
\nonumber \\
&\times\exp\left[-\beta\tilde{H}^{\rm cl}(\{J_p+1,\theta_p,\phi_p\})\right].
\end{align*}
Here ${\cal W}(\{J_p\})$ is defined by ${\cal W}(\{J_p\})\equiv\prod_pW(J_p)$,
and $D\Omega\equiv d\Omega_1d\Omega_2\dots d\Omega_{(L/l)^d}$.

\section{Variational expression of the free energy}
\label{sec:variational}

In this section, we evaluate the upper and the lower bounds of the free energy and derive its variational expression.
Let us define
\beq
\inf_{\{\theta_p,\phi_p\}}\tilde{H}^{\rm cl}(\{ J_p,\theta_p,\phi_p\})\equiv E^*(\{ J_p\}).
\eeq
We then find
\beq
Z_+\leq\sum_{\{ J_p\}}{\cal W}(\{ J_p\})\left[\prod_p^{(L/l)^d}(2J_p+1)\right]e^{-\beta E^*(\{ J_p+1\})}.
\eeq
Because $\sum_{\{ J_p\}}f(\{ J_p\})\leq (2sl^d)^{(L/l)^d}\max_{\{ J_p\}}f(\{ J_p\})$ for an arbitrary function $f(\{ J_p\})$,
\begin{align}
Z_+\leq (2sl^d)^{(L/l)^d}\exp\left\{\max_{\{ J_p\}}
\Bigg[-\beta E^*(\{ J_p+1\})\right.
\nonumber \\
+\sum_p^{(L/l)^d}\ln(2J_p+1)+\ln {\cal W}(\{ J_p\})\Bigg]\Bigg\}.
\end{align}
Therefore, we obtain
\begin{align}
{\rm Lim} f\geq {\rm Lim}\inf_{\{\vec{S}_p\}}\frac{1}{L^d}\bigg[ \tilde{H}^{\rm cl}(\{ J_p+1,\theta_p,\phi_p\})
\nonumber \\
-\frac{1}{\beta}\ln {\cal W}(\{ J_p\})\bigg].
\label{eq:lower1}
\end{align}

We show that we can replace $J_p+1$ by $J_p$ in Eq.~(\ref{eq:lower1}).
It is verified by the following evaluation:
\begin{align}
\left|\tilde{H}^{\rm cl}(\{ J_p+1,\theta_p,\phi_p\})-\tilde{H}^{\rm cl}(\{ J_p,\theta_p,\phi_p\})\right|
\nonumber \\
=\frac{J_p+J_q+1}{2s^2}\left|\sum_{pq}^{(L/l)^d}\gamma^d\phi_{pq}\sum_{a=x,y,z}\lambda_{\alpha}(\vec{e}_p)^a(\vec{e}_q)^a\right|,
\end{align}
where $\vec{e}_p=\vec{S}_p/|\vec{S}_p|$.
As $J_p\leq sl^d$, we have
\begin{align}
\left|\tilde{H}^{\rm cl}(\{ J_p+1,\theta_p,\phi_p\})-\tilde{H}^{\rm cl}(\{ J_p,\theta_p,\phi_p\})\right|
\nonumber \\
\leq \frac{2sl^d+1}{2s^2}\lambda_{\rm max}\sum_{pq}^{(L/l)^d}\gamma^d\phi_{pq}.
\end{align}
Here we note that
\beq
\sum_{pq}^{(L/l)^d}\gamma^d\phi_{pq}=\frac{1}{l^{2d}}\sum_{ij}\gamma^d\phi(\gamma\bm{r}_{ij})=\frac{L^d}{l^{2d}},
\eeq
and obtain
\begin{align}
{\rm Lim}\frac{1}{L^d}\left|\tilde{H}^{\rm cl}(\{ J_p+1,\theta_p,\phi_p\})-\tilde{H}^{\rm cl}(\{ J_p,\theta_p,\phi_p\})\right|
\nonumber \\
\leq {\rm Lim}\frac{2sl^d+1}{2s^2l^{2d}}\lambda_{\rm max}=0.
\end{align}
Thus we obtain the lower bound
\beq
{\rm Lim}f\geq {\rm Lim}\inf_{\{\vec{S}_p\}}\frac{1}{L^d}\left[\tilde{H}^{\rm cl}(\{ J_p,\theta_p,\phi_p\})-\frac{1}{\beta}\ln {\cal W}(\{ J_p\})\right].
\label{eq:f_lower}
\eeq

Next we estimate the upper bound.
We define
\beq
(\theta_p^*,\phi_p^*)\equiv\arginf_{\{\theta_p,\phi_p\}}\tilde{H}^{\rm cl}(\{ J_p,\theta_p,\phi_p\}),
\eeq
$\theta_p'\equiv|\theta_p-\theta_p^*|$, and $\phi_p'\equiv{\rm mod}(\phi_p-\phi_p^*,2\pi)$.
In addition, the set $\omega_p(\Delta)$ is defined by
\beq
\omega_p(\Delta)\equiv\{(\theta_p,\phi_p):\theta_p'\in[0,\Delta]\}.
\eeq
For sufficiently small $\Delta$, the volume of the set $\omega_p(\Delta)$ satisfies
\beq
\int_{(\theta_p,\phi_p)\in\omega_p(\Delta)}d\Omega_p=2\pi(1-\cos\Delta)
>\Delta^2.
\label{eq:area_omega_p}
\eeq
Moreover, we can show that
\beq
\sup_{(\theta_p,\phi_p)\in\omega_p(\Delta)}\tilde{H}^{\rm cl}(\{ J_p,\theta_p,\phi_p\})-E^*(\{ J_p\})
\leq L^d\lambda_{\rm max}\Delta,
\label{eq:Delta}
\eeq
whose proof is given in Appendix~\ref{sec:Delta}.

From Eqs.~(\ref{eq:area_omega_p}) and (\ref{eq:Delta}),
\begin{align}
Z_-\geq&\sum_{\{ J_p\}}{\cal W}(\{ J_p\})\frac{1}{(4\pi)^{(L/l)^d}}\int_{\{(\theta_p,\omega_p)\in\omega_p(\Delta)\}}D\Omega
\nonumber \\
&\times\exp\left[-\beta\tilde{H}^{\rm cl}(\{ J_p,\theta_p,\phi_p\})\right] \nonumber \\
\geq&\sum_{\{ J_p\}}{\cal W}(\{ J_p\})\frac{1}{(4\pi)^{(L/l)^d}}
\nonumber \\
&\times\exp\left[-\beta\sup_{\{(\theta_p,\phi_p)\in\omega_p(\Delta)\}}\tilde{H}^{\rm cl}(\{ J_p,\theta_p,\phi_p\})\right]
\nonumber \\
&\times\left[\int_{(\theta_p,\phi_p)\in\omega_p(\Delta)}d\Omega_p\right]^{(L/l)^d} \nonumber \\
\geq&\sum_{\{ J_p\}}{\cal W}(\{ J_p\})\left(\frac{\Delta^2}{4\pi}\right)^{(L/l)^d}
\nonumber \\
&\times\exp\left[-\beta E^*(\{ J_p\})-\beta L^d\lambda_{\rm max}\Delta\right] \nonumber \\
\geq&\max_{\{ J_p\}}\exp\Bigg[-\beta E^*(\{ J_p\})+\ln {\cal W}(\{ J_p\})
\nonumber \\
&+\left(\frac{L}{l}\right)^d\ln\frac{\Delta^2}{4\pi}
-\beta L^d\lambda_{\rm max}\Delta\Bigg].
\end{align}
Since $\Delta>0$ is arbitrary, we choose $\Delta=l^{-d}$.
We then find
\beq
{\rm Lim}f\leq{\rm Lim}\inf_{\{\vec{S}_p\}}\frac{1}{L^d}\left[\tilde{H}^{\rm cl}(\{ J_p,\theta_p,\phi_p\})-\frac{1}{\beta}\ln {\cal W}(\{ J_p\})\right].
\label{eq:f_upper}
\eeq

From the upper bound~(\ref{eq:f_upper}) and the lower bound~(\ref{eq:f_lower}), we obtain the desired variational expression for the free energy,
\beq
{\rm Lim}f={\rm Lim}\inf_{\{\vec{S}_p\}}\frac{1}{L^d}\left[\tilde{H}^{\rm cl}(\{ J_p,\theta_p,\phi_p\})-\frac{1}{\beta}\ln {\cal W}(\{ J_p\})\right].
\label{eq:f_variational}
\eeq
The reason why we can reach the variational expression~(\ref{eq:f_variational}) is 
that the degrees of freedom to be considered is dramatically decreased by the coarse graining.

We can express the free energy by the continuous functional.
If we see the system in the length scale so the size of the system is unity,
\begin{align}
f=\inf_{\vec{S}(\bm{x})}&\left[-\frac{1}{2}\int_{[0,1)^d}d^dx\int_{[0,1)^d}d^dy U(\bm{x}-\bm{y})\right.
\nonumber \\
&\times\sum_{a=x,y,z}\lambda_aS^a(\bm{x})S^a(\bm{y})
\nonumber \\
&\left. -\vec{h}\cdot\int_{[0,1)^d}\vec{S}(\bm{x})d^dx
-\frac{1}{\beta}\int_{[0,1)^d}\sigma(\vec{S}(\bm{x}))d^dx\right] \nonumber \\
\equiv\inf_{\vec{S}(\bm{x})}&{\cal F}[\vec{S}(\bm{x})].
\label{eq:f_continuum}
\end{align}
Here we defined the scaled potential
$$U(\bm{x})\equiv{\rm Lim}(\gamma L)^d\phi(\gamma L\bm{x}),$$
and the entropy function
$$\sigma(\vec{S}(\bm{x}))\equiv{\rm Lim}\frac{1}{l^d}\ln W(J_p),$$
with $\vec{S}(\bm{x})\equiv \vec{S}_p/(sl^d)$ and $\bm{x}\equiv{\rm Lim}(\bm{r}_p/L)$, where $\bm{r}_p$ is the central position of the cell ${\rm C}_p$.
Notice that the length of the spin vector does not exceed unity, $|\vec{S}(\bm{x})|\leq 1$.

The {\it free energy functional} ${\cal F}[\vec{S}(\bm{x})]$ consists of two parts, 
$${\cal F}[\vec{S}(\bm{x})]={\cal U}[\vec{S}(\bm{x})]-\frac{1}{\beta}\int_{[0,1]^d}\sigma(\vec{S}(\bm{x}))d^dx,$$ where
\begin{align}
{\cal U}[\vec{S}]=-\frac{1}{2}\int_{[0,1)^d}d^dx\int_{[0,1)^d}d^dy U(\bm{x}-\bm{y})
\nonumber \\
\times\sum_{a=x,y,z}\lambda_aS^a(\bm{x})S^a(\bm{y})
-\vec{h}\cdot\int_{[0,1)^d}\vec{S}(\bm{x})d^dx
\label{eq:internal}
\end{align}
is the {\it internal energy functional}.
It corresponds to the energy of a given configuration $\vec{S}(\bm{x}))$.
$\int\sigma(\vec{S}(\bm{x})d^dx$ corresponds to the entropy of a given configuration $\vec{S}(\bm{x})$.
Because many microscopic states correspond to a single coarse-grained configuration $\vec{S}(\bm{x})$, this entropic term arises.
Thus, the calculation of the free energy is reduced to the minimization problem of the free energy functional.

It is noted that if the uniform spin configuration, $\vec{S}(\bm{x})=\vec{m}$ independent of $\bm{x}$, is assumed,
the free energy is exactly equal to that of the mean-field model (the model with $\phi(\bm{x})=1/N$) and independent of the potential $\phi(\bm{x})$;
\beq
{\cal F}[\vec{S}(\bm{x})=\vec{m}]=f_{\rm MF}(\beta,\vec{m},\vec{h}).
\eeq
The explicit form of $f_{\rm MF}(\beta,\vec{m},\vec{h})$ is given in Eq.~(\ref{eq:MF_free}).
This observation implies that we can replace the interaction potential $\phi(\bm{x})$ by the simple infinite range potential $1/N$
{\it as long as the magnetization profile is uniform in equilibrium}.
While, the spin configuration must be inhomogeneous whenever this replacement is not allowed.

Finally, I make a few remarks on the relation to previous works.
Gates and Penrose~\cite{Gates1969,Gates1970} obtained the similar variational expression of the free energy.
However, the free energy functional obtained in the present paper [Eq.~(\ref{eq:f_continuum})] 
differs from that in those works with respect to the length scale of the free energy functional.
In the present work, the position variable $\bm{x}$ runs over $d$-dimensional unit cube, $\bm{x}\in[0,1]^d$, and the system size is set to be unity.
It enables us to treat the long-range interaction potential like $1/r^{\alpha}$, $\alpha<d$.
On the other hand, in the free energy functional obtained by Ref.~\cite{Gates1969},
the position variable $\bm{x}$ runs over the whole $d$-dimensional space, $\bm{x}\in\mathbb{R}^d$.
In that case, we cannot treat the long-range interaction like $1/r^{\alpha}$ with $\alpha<d$ because the integrated value of the potential is divergent.
If we consider only the van der Waals limit, however, 
the free energy functional in Ref.~\cite{Gates1969} has an advantage that it has information on the mesoscopic length scale.
In the expression~(\ref{eq:f_continuum}), we cannot obtain any information on the length scale $\sim \gamma^{-1}$, 
{\it e.g.} the form of the interface in an ordered state with phase separation.

In the work by Kiessling and Percus~\cite{Kiessling1995}, 
the conceptually different but mathematically equivalent limiting procedure was introduced to study the liquid-vapor interface in a many particle classical system.
That is, they took the limit of infinitely many particles in a finite domain.
As a result, they derived the similar variational expression as Eq.~(\ref{eq:f_continuum}) on the thermodynamic potential.
It shows the similarity between bulk properties of a system with a nonadditive long-range interaction and
local properties of a system with a finite-range interaction and with a very large number of particles in a bounded domain.
In the works of Refs.~\cite{Mori2010_analysis,Mori2011_instability,Mori2012_microcanonical}, from the variational expression~(\ref{eq:f_continuum}),
simple sufficient and necessary conditions for the exactness of the mean-field theory are derived
in classical spin systems, which is presented in the next section.

\section{Exactness of the mean-field theory}
\label{sec:exactness}

Since the calculation of the free energy is reduced to the minimization of the free energy functional expressed by the classical variables,
we can apply the same argument as the case of classical spin systems.
In this section, we briefly review the result of the variational expression~(\ref{eq:f_variational}) or equivalently, the continuum expression~(\ref{eq:f_continuum}).

Hereafter, we assume periodic boundary conditions and apply the nearest image convention to pair interactions.

\subsection{Canonical ensemble}
\label{sec:canonical}
We assume that the system is in contact with a thermal reservoir at a temperature $T=1/\beta$ and there is no conserved quantity.
In this case, the free energy is independent of the interaction potential $\phi(\cdot)$ and it is equal to the mean-field free energy
\begin{align}
f_{\rm MF}(\beta)&=\min_{\vec{m}}\left[-\frac{1}{2}\sum_{a=x,y,z}\lambda_am_a^2-\vec{h}\cdot\vec{m}-\frac{1}{\beta}\sigma(\vec{m})\right]
\nonumber \\
&\equiv\min_{\vec{m}}f_{\rm MF}(\beta,\vec{m},\vec{h}),
\label{eq:MF_free}
\end{align}
where
$$\vec{m}=\int_{[0,1)^d}\vec{S}(\bm{x})d^dx.$$
This result is called ``exactness of the mean-field theory'' in long-range interacting systems.

It is convenient to consider the canonical ensemble with the fixed magnetization, although the magnetization $\vec{m}$ is not a conserved quantity.
Roughly speaking, the free energy
\beq
f(\beta,\vec{m},\vec{h})=\min_{\{\vec{S}(\bm{x}):\vec{m}=\int_{[0,1)^d}\vec{S}(\bm{x})d^dx\}}{\cal F}[\vec{S}(\bm{x})]
\eeq
is related to the probability $P(\vec{m})$ of the states with the magnetization $\vec{m}$ by $P(\vec{m})\sim\exp[-\beta f(\beta,\vec{m},\vec{h})]$,
although the meaning of ``the states with the magnetization $\vec{m}$'' is ambiguous in quantum spin systems because of the quantum fluctuation.
This function is well-defined and  indeed useful when we analyze the case 
where there is some conserved quantity like the energy in Sec.~\ref{sec:micro} or $m_z$ in the XXZ Heisenberg model discussed in Sec.~\ref{sec:ex}.

Later, we write $f(\beta,\vec{m},\vec{h})$ simply as $f(\beta,\vec{m})$ because the dependence on $\vec{h}$ is trivial.

In the van der Waals limit, the scaled potential is the $\delta$ function, $U(\bm{x})=\delta(\bm{x})$, and in this case we can show
\beq
f(\beta,\vec{m})=f_{\rm MF}^{**}(\beta,\vec{m}).
\eeq
This corresponds to the Lebowitz-Penrose theorem.

In the nonadditive limit, $U(\bm{x})=\phi(\bm{x})$.
The Fourier expansion is given by
\beq
U(\bm{x})=\sum_{\bm{n}\in\mathbb{Z}^d}U_{\bm{n}}e^{2\pi i\bm{n}\cdot\bm{x}}.
\eeq
If we define 
\beq
U_{\max}\equiv\max_{\bm{n}\in\mathbb{Z}^d\backslash 0}U_{\bm{n}},
\eeq
we obtain~\cite{Mori2010_analysis,Mori2011_instability}
\begin{align}
f_{\rm MF}(\beta,\vec{m})-U_{\rm max}\Delta f_{\rm MF}(\beta U_{\rm max},\vec{m})
\nonumber \\
\leq f(\beta,\vec{m})\leq f_{\rm MF}(\beta,\vec{m}),
\label{eq:f_ineq}
\end{align}
where $\Delta f_{\rm MF}\equiv f_{\rm MF}-f_{\rm MF}^{**}$,
and moreover
\beq
f(\beta,\vec{m})<f_{\rm MF}(\beta,\vec{m})
\eeq 
when the matrix
\beq
[{\cal L}(\beta U_{\rm max},\vec{m})]_{ab}\equiv\frac{\d^2f_{\rm MF}}{\d m_a\d m_b}(\beta U_{\rm max},\vec{m})
\label{eq:L}
\eeq
is non-positive.

From the inequality~(\ref{eq:f_ineq}), we find that $f(\beta,\vec{m})=f_{\rm MF}(\beta,\vec{m})$ in the parameter $(\beta,\vec{m})$ such that
$f_{\rm MF}(\beta U_{\rm max},\vec{m})=f_{\rm MF}^{**}(\beta U_{\rm max},\vec{m})$.

We define $\tilde{\beta}$ by
\beq
\tilde{\beta}\equiv\sup\left[\beta>0:f_{\rm MF}(\beta,\vec{m})=f_{\rm MF}^{**}(\beta,\vec{m}) \quad \forall \vec{m}\right].
\label{eq:tilde_beta}
\eeq
It is then concluded that $f(\beta,\vec{m})=f_{\rm MF}(\beta,\vec{m})$ for any $\vec{m}$ when $\beta U_{\rm max}<\tilde{\beta}$.
This $\tilde{\beta}$ plays important roles in the following sections.

\subsection{Microcanonical ensemble}
\label{sec:micro}

In this section we consider the microcanonical ensemble without any other conserved quantities.
The case in which there is another conserved quantity like $m_z$ is treated in Sec.~\ref{sec:ex}.
The microcanonical entropy is obtained by
\beq
s(\varepsilon)=\sup_{\{\vec{S}(\bm{x})\}}\left[\int_{[0,1)^d}\sigma(\vec{S}(\bm{x}))d^dx:\varepsilon={\cal U}[\vec{S}(\bm{x})]\right],
\eeq
where the internal energy functional ${\cal U}[\vec{S}(\bm{x})]$ is defined by Eq.~(\ref{eq:internal}).
This expression is common to classical spin systems, so the same argument as classical spin systems~\cite{Mori2012_microcanonical} can be applied.

We can easily verify
\beq
s_{\rm MF}(\varepsilon)\leq s(\varepsilon)\leq s_{\rm MF}^{**}(\varepsilon),
\label{eq:entropy_ineq}
\eeq
where $s_{\rm MF}$ is the microcanonical entropy in the mean-field model, $\phi(\bm{x})=1$,
and $s_{\rm MF}^{**}(\varepsilon)$ is the concave envelope of $s_{\rm MF}(\varepsilon)$ or 
the function obtained by performing the Legendre transformation twice on $s_{\rm MF}(\varepsilon)$.
From the inequality~(\ref{eq:entropy_ineq}), we can conclude that 
the mean-field theory is exact in the microcanonical ensemble, $s(\varepsilon)=s_{\rm MF}(\varepsilon)$,
if the canonical and the microcanonical ensembles are equivalent in the mean-field model because $s_{\rm MF}=s_{\rm MF}^{**}$ in this case.

In the case where the canonical and the microcanonical ensembles are inequivalent in the mean-field model, more sophisticated treatment is necessary.
In this case, it was shown in~\cite{Mori2012_microcanonical} that
\begin{align}
&s_{\rm MF}(\varepsilon)\leq s(\varepsilon)
\nonumber \\
&\leq
\min\left\{ s_{\rm MF}^{**}(\varepsilon), 
\sup_{\varepsilon'>\varepsilon}\left[s_{\rm MF}(\varepsilon')+\frac{\tilde{\beta}}{U_{\rm max}}(\varepsilon-\varepsilon')\right]\right\}.
\label{eq:s_inequality2}
\end{align}
Here $\tilde{\beta}$ is given by Eq.~(\ref{eq:tilde_beta}).
Moreover, it was shown that 
\beq
s(\varepsilon)>s_{\rm MF}(\varepsilon)
\label{eq:s_instability}
\eeq
when the matrix
${\cal L}(\beta_{\rm MF}(\varepsilon)U_{\rm max},\vec{m}_{\rm eq}(\varepsilon))$ is nonpositive.
The definition of the matrix ${\cal L}$ is given by Eq.~(\ref{eq:L}).
The quantity $\beta_{\rm MF}(\varepsilon)=\d s_{\rm MF}/\d\varepsilon$ is the inverse temperature of the mean-field model at the energy $\varepsilon$,
and $\vec{m}_{\rm eq}(\varepsilon)$ is the equilibrium magnetization of the mean-field model,
\beq
\vec{m}_{\rm eq}(\varepsilon)\equiv\argsup_{\vec{m}}
\left[\sigma(\vec{m}):\varepsilon=-\frac{1}{2}\sum_{a=x,y,z}\lambda_am_a^2-\vec{h}\cdot\vec{m}\right].
\eeq

The result of the inequality~(\ref{eq:s_inequality2}) is as follows.
We consider the case that $s_{\rm MF}(\varepsilon)<s_{\rm MF}^{**}(\varepsilon)$ in the region of $\varepsilon_a<\varepsilon<\varepsilon_b$.
We define
\beq
\beta^*\equiv\sup\left[\frac{\d s_{\rm MF}(\varepsilon)}{\d\varepsilon}:\varepsilon\in (\varepsilon_a,\varepsilon_b)\right].
\eeq
If $\beta^*U_{\rm max}<\tilde{\beta}$, then $s(\varepsilon)=s_{\rm MF}(\varepsilon)$ holds for all the values of $\varepsilon$.
Therefore, the states with the negative specific heats in the mean-field model are maintained for systems with general long-range interaction $\phi(\cdot)$
as long as the condition $\beta^*U_{\rm max}<\tilde{\beta}$ is satisfied.

On the other hand, the inequality~(\ref{eq:s_instability}) means that when the matrix
${\cal L}(\beta_{\rm MF}(\varepsilon)U_{\rm max},\vec{m}_{\rm eq}(\varepsilon))$ is non-positive,
the equilibrium state in the mean-field model, $\vec{S}(\bm{x})=\vec{m}_{\rm eq}(\varepsilon)$, becomes unstable and the inhomogeneity appears.
In this case, the mean-field theory is not exact.

\section{Example: the spin-1/2 XXZ model}
\label{sec:ex}

As an application of the theoretical framework, we consider the spin-1/2 XXZ model, 
whose Hamiltonian is given by Eq.~(\ref{eq:H}) with $\lambda_x=\lambda_y\equiv\lambda_{\perp}\geq 0$ and $\lambda_z\geq 0$.
The Hamiltonian is given by
\beq
H=-\frac{1}{2}\sum_{ij}\gamma^d\phi(\gamma\bm{r}_{ij})\left[\lambda_{\perp}(s_i^xs_j^x+s_i^ys_j^y)+\lambda_zs_i^zs_j^z\right].
\label{eq:H_XXZ}
\eeq
The mean-field counterpart of this model [$\gamma^d\phi(\gamma\bm{r}_{ij})=1/N$ in Eq.~(\ref{eq:H_XXZ})] 
was studied by M. Kastner~\cite{Kastner2010,Kastner_statmech2010}.

When the magnetization is not fixed,
the exactness of the mean-field theory always holds in the canonical ensemble, $f(\beta,\vec{h})=f_{\rm MF}(\beta,\vec{h})$,
which is a general result from Sec.~\ref{sec:canonical}.
Moreover, in this model, the mean-field theory is exact even in the microcanonical ensemble, $s(\varepsilon)=s_{\rm MF}(\varepsilon)$.
It is resulted from the fact that 
the microcanonical entropy of the mean-field model corresponding to Eq.~(\ref{eq:H_XXZ}) is a concave function of $\varepsilon$.
It implies that $s_{\rm MF}(\varepsilon)=s_{\rm MF}^{**}(\varepsilon)$ and, thus, $s(\varepsilon)=s_{\rm MF}(\varepsilon)$ from Eq.~(\ref{eq:entropy_ineq}).

If the system is isolated, the $z$ component of the magnetization $m_z$ is conserved.
In this section, we consider the canonical and the microcanonical ensembles with a fixed value of the $z$ component of the magnetization $m_z$.
In this case, as we will see below, the result is nontrivial and the exactness of the mean-field theory is violated for certain parameter region.

\subsection{Canonical analysis}

First, we investigate the model in the canonical ensemble.
The free energy as a function of $\beta$ and $m_z$ is given by
\beq
f(\beta,m_z)=\inf_{\{ m_x,m_y\}}f(\beta,\vec{m})\equiv f(\beta,\vec{m}_{\rm eq}(\beta,m_z)).
\eeq
Of course, $(\vec{m}_{\rm eq}(\beta,m_z))_z=m_z$.

The analysis of the previous sections leads us to the following result.
From the inequality~(\ref{eq:f_ineq}), we obtain
\begin{align}
f_{\rm MF}(\beta,\vec{m}_{\rm eq}(\beta,m_z))&-U_{\rm max}\Delta f_{\rm MF}(\beta U_{\rm max},\vec{m}_{\rm eq}(\beta,m_z))
\nonumber \\
&\leq f(\beta,m_z)\leq f_{\rm MF}(\beta,m_z).
\end{align}
Therefore, if $\beta U_{\rm max}<\tilde{\beta}$, $f(\beta,m_z)=f_{\rm MF}(\beta,m_z)$ holds because $\Delta f_{\rm MF}(\beta U_{\rm max},\vec{m})=0$.
Moreover, if the matrix ${\cal L}(\beta U_{\rm max},\vec{m}_{\rm eq}(\beta,m_z))$ has negative eigenvalues,
the mean-field model is not exact, $f(\beta,m_z)<f_{\rm MF}(\beta,m_z)$.

Since the entropy function of the spin-1/2 system is given by Eq.~(\ref{eq:entropy_spin1/2}) in Appendix~\ref{sec:W},
the mean-field free energy with a fixed $\vec{m}$ is given by
\begin{align}
f_{\rm MF}(\beta,\vec{m})=-\frac{1}{2}\left[\lambda_{\perp}(m_x^2+m_y^2)+\lambda_zm_z^2\right]
\nonumber \\
+\frac{1}{\beta}\left(\frac{1+|\vec{m}|}{2}\ln\frac{1+|\vec{m}|}{2}+\frac{1-|\vec{m}|}{2}\ln\frac{1-|\vec{m}}{2}\right).
\label{eq:f_XXZ}
\end{align}
In order to obtain the mean-field free energy as a function of $m_z$, it is necessary to minimize $f_{\rm MF}(\beta,\vec{m})$ for $m_x$ and $m_y$.
By doing that, we find in the mean-field model
\beq
\left\{
\begin{split}
&(\vec{m}_{\rm eq})_x=(\vec{m}_{\rm eq})_y=0 \text{ when } m_z\geq\tanh(\beta\lambda_{\perp}m_z), \\
&|\vec{m}_{\rm eq}|=\tanh(\beta\lambda_{\perp}|\vec{m}_{\rm eq}|) \text{ when }m_z<\tanh(\beta\lambda_{\perp}m_z).
\end{split}
\right.
\eeq
By differentiating twice the mean-field free energy $f_{\rm MF}(\beta,\vec{m})$, we obtain the Hessian matrix ${\cal L}(\beta,\vec{m})$ (see Eq.~(\ref{eq:L})).
We can obtain $\tilde{\beta}$, which plays important roles to judge whether the mean-field theory is exact at a given parameter $(\beta,m_z)$,
by analyzing the eigenvalues of ${\cal L}(\beta,\vec{m})$.
Actually, we obtain 
\beq
\tilde{\beta}=\min\left\{\lambda_{\perp}^{-1},\lambda_z^{-1}\right\}.
\label{eq:tilde_XXZ}
\eeq
See Appendix~\ref{sec:tilde_XXZ} for the proof of Eq.~(\ref{eq:tilde_XXZ}).

By the reason mentioned in later analysis on the microcanonical ensemble, $f(\beta,m_z)=f_{\rm MF}(\beta,m_z)$ when $\lambda_{\perp}>\lambda_z$.
The result is non-trivial when $\lambda_z>\lambda_{\perp}$.
In this case, $\tilde{\beta}=1/\lambda_z$, and the mean-field model gives the exact free energy, $f(\beta,m_z)=f_{\rm MF}(\beta,m_z)$
at least for $\beta<1/(U_{\rm max}\lambda_z)$.

By calculating the minimum eigenvalue of ${\cal L}(\beta U_{\rm max},\vec{m}_{\rm eq}(\beta))$,
we obtain the sufficient condition under which the exactness of the mean-field theory does {\it not} hold.

\begin{figure*}[tbh]
\begin{center}
\begin{tabular}{ccc}
(a)&(b)&(c)\\
\includegraphics[clip,width=4cm]{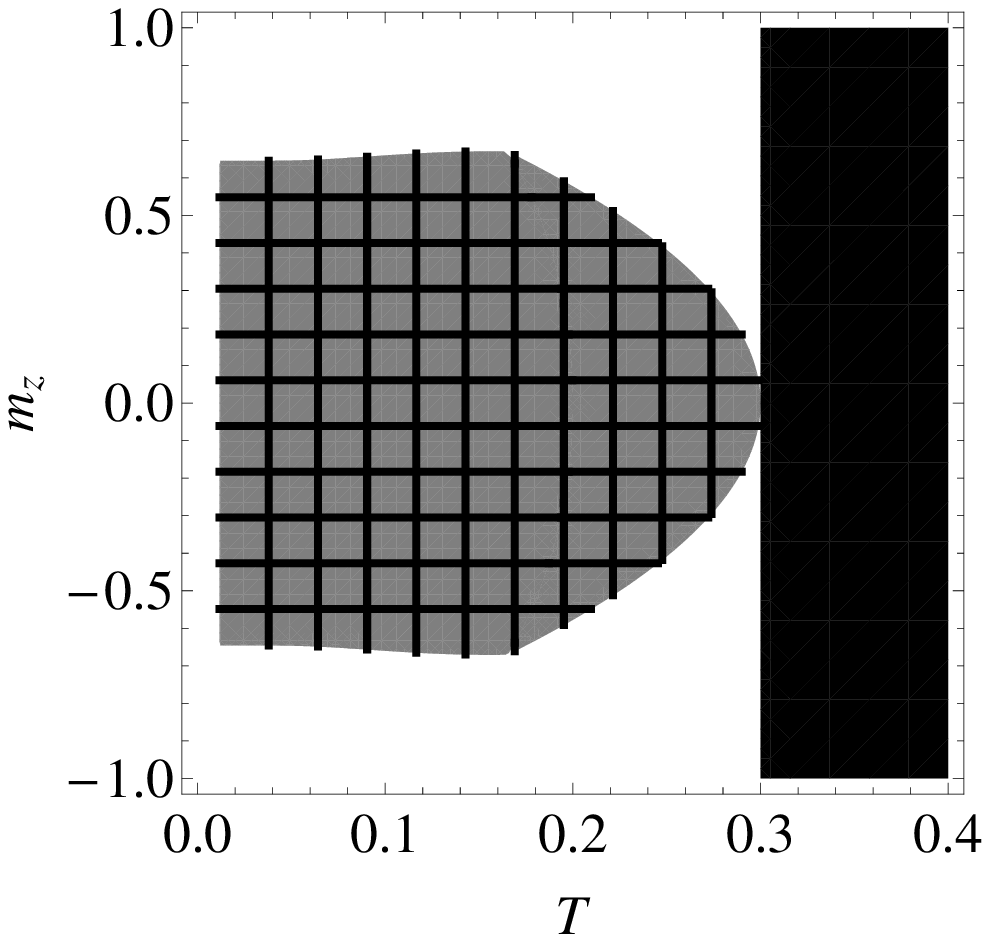}&
\includegraphics[clip,width=4cm]{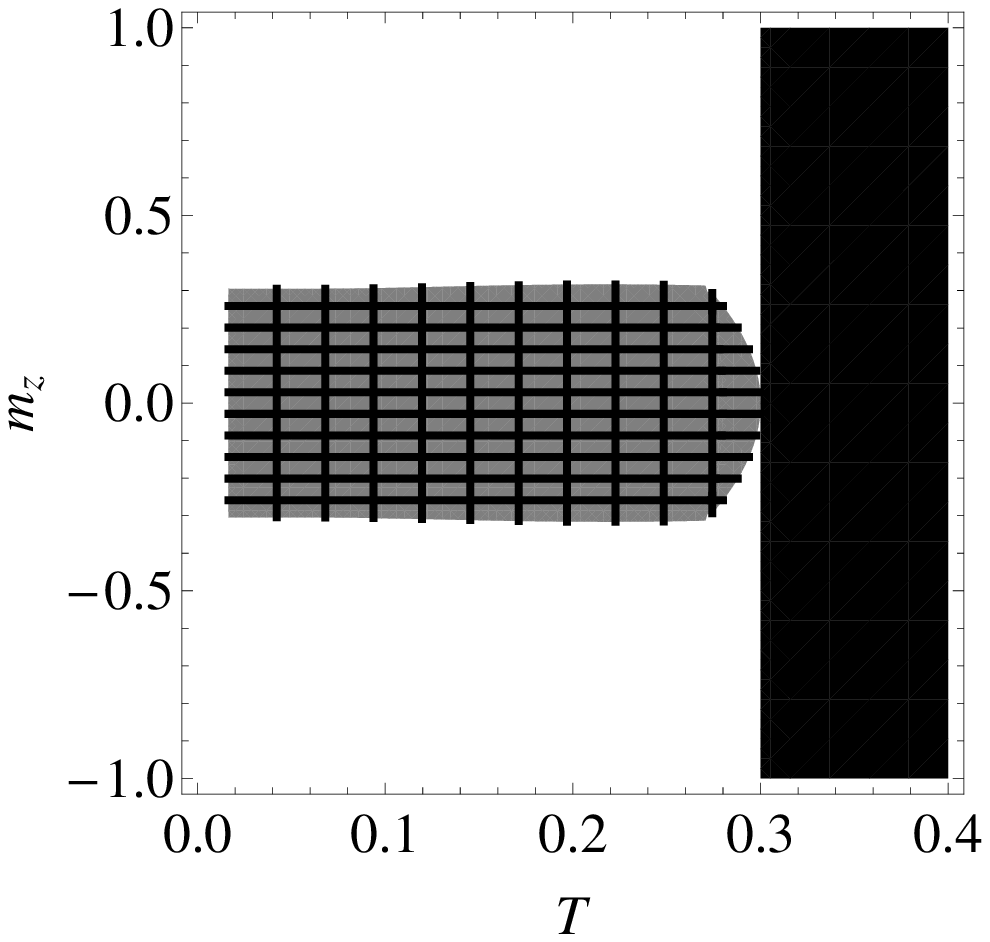}&
\includegraphics[clip,width=4cm]{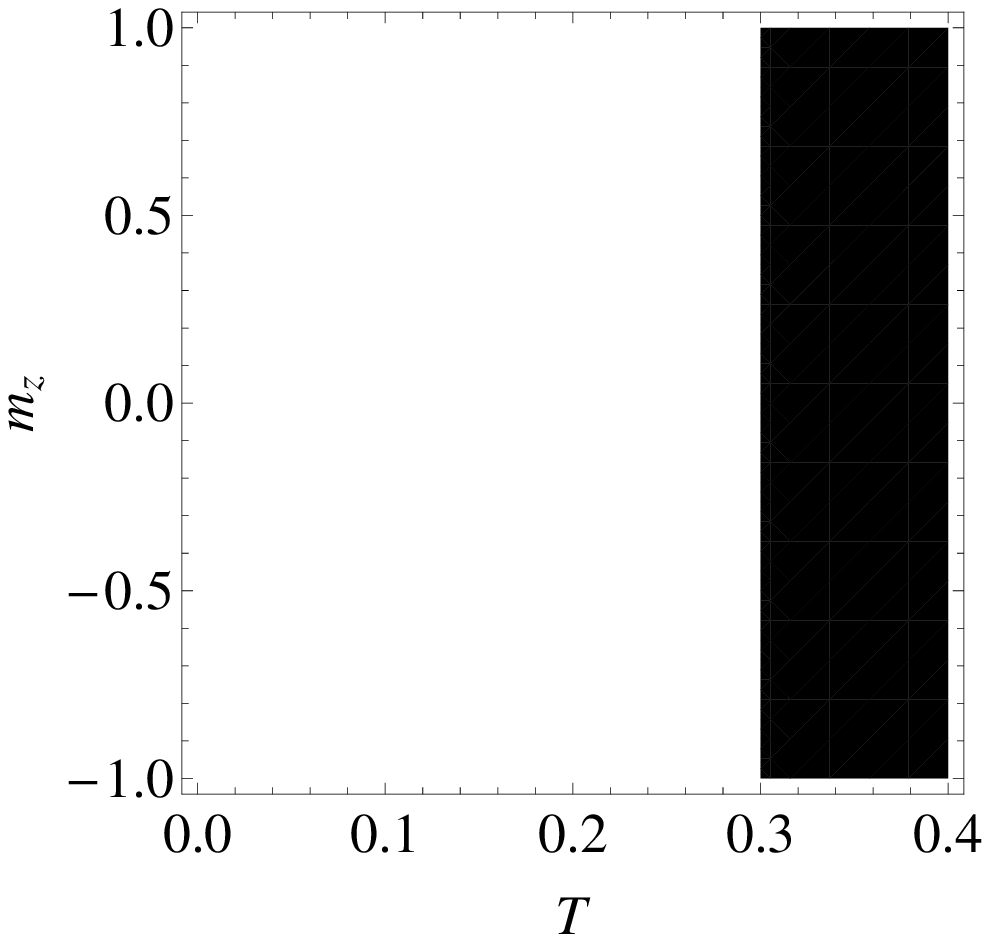}
\end{tabular}
\caption{The abscissa is the temperature $T=1/\beta$ and the ordinate is $m_z$.
In the black region, the mean-field theory is exact, and in the meshed gray region, the mean-field theory is not exact.
We cannot determine whether the mean-field theory is exact in the white region.
The parameters $\lambda_z$ and $U_{\rm max}$ are set to 1 and 0.3, respectively.
(a) $\lambda_{\perp}=0.2$, (b) $\lambda_{\perp}=0.28$, and (c) $\lambda_{\perp}=0.4$. }
\label{fig:XXZ_canonical}
\end{center}
\end{figure*}

In Fig.~\ref{fig:XXZ_canonical}, we plot the two regions in $(T=1/\beta,m_z)$ plane.
One is the region which is given by $\beta<1/(U_{\rm max}\lambda_z)$ (black region).
In this region, the mean-field theory is exact. 
The other is the region where the exactness of the mean-field theory is violated (meshed gray region),
which is given by the condition that the minimum eigenvalue of ${\cal L}(\beta U_{\rm max},\vec{m}_{\rm eq}(\beta))$ is negative.
The parameter $U_{\rm max}$ is set to be 0.3.
This value corresponds to the case that $\phi(\bm{r})\propto 1/r$ in the two-dimensional lattice.
In the white region in Fig.~\ref{fig:XXZ_canonical}, we cannot determine whether the mean-field theory is exact from our analysis.
However, in the white region, the equilibrium state predicted by the mean-field model is at least locally stable 
because all the eigenvalues of ${\cal L}(\beta U_{\rm max},\vec{m}_{\rm eq}(\beta))$ are positive.

\subsection{Microcanonical analysis}

Next, we consider the model in the microcanonical ensemble.
The mean-field entropy was obtained and the property of the mean-field model was studied extensively~\cite{Kastner2010,Kastner_statmech2010}.
Some of the results obtained here will overlap these studies.
The microcanonical entropy in the mean-field model is given by
\beq
s_{\rm MF}(\varepsilon,m_z)=-\frac{1+g}{2}\ln\frac{1+g}{2}-\frac{1-g}{2}\ln\frac{1-g}{2},
\label{eq:s_XXZ}
\eeq
where
\beq
g=\sqrt{m_z^2(1-\lambda_z/\lambda_{\perp})-2\varepsilon/\lambda_{\perp}}.
\eeq
Because there is the conserved quantity $m_z$ besides the energy $\varepsilon$,
the domain of the entropy is not necessarily a convex set.
The domain is given by
\begin{align}
{\cal D}_{\rm MF}=&\left\{(\varepsilon,m_z):\varepsilon<-\frac{\lambda_z}{2}m_z^2\right.
\nonumber \\
&\left.\text{ and }
\varepsilon>\frac{\lambda_{\perp}-\lambda_z}{2}m_z^2-\frac{1}{2}\lambda_{\perp}\right\}.
\end{align}
The domain of the model~(\ref{eq:H_XXZ}), which is referred to as ${\cal D}$, and that of the mean-field model, ${\cal D}_{\rm MF}$, differ in general,
$${\cal D}\supset{\cal D}_{\rm MF}.$$
Exactness of the mean-field theory is trivially violated for parameters $(\varepsilon,m_z)\in{\cal D}\backslash{\cal D}_{\rm MF}$.

When $m_z$ is conserved, the analysis given in Sec.~\ref{sec:micro} must be slightly modified.
Inequality~(\ref{eq:s_inequality2}) is altered to
\begin{align}
&s_{\rm MF}(\varepsilon,m_z)\leq s(\varepsilon,m_z)
\nonumber \\
&\leq\min\left\{ 
s_{\rm MF}^{**}(\varepsilon,m_z),
\sup_{\varepsilon'>\varepsilon}\left[s_{\rm MF}(\varepsilon',m_z)+\frac{\tilde{\beta}}{U_{\rm max}}(\varepsilon-\varepsilon')\right]\right\},
\label{eq:s_inequality2-2}
\end{align}
where $s_{\rm MF}^{**}(\varepsilon,m_z)$ is the function obtained by applying twice the Legendre transformation to $s_{\rm MF}(\varepsilon,m_z)$ 
{\it with respect to both variables} $(\varepsilon,m_z)$\footnote
{This is the same as the minimum concave function of $\varepsilon$ and $m_z$ which is not less than $s_{\rm MF}(\varepsilon,m_z)$.}.
From the inequality~(\ref{eq:s_inequality2-2}), it is concluded that $s(\varepsilon,m_z)=s_{\rm MF}(\varepsilon,m_z)$ 
at least when $(\varepsilon,m_z)\in{\cal D}_{\rm MF}$ and $\beta_{\rm MF}(\varepsilon,m_z)<\tilde{\beta}/U_{\rm max}$.
Here, $\beta_{\rm MF}(\varepsilon,m_z)\equiv\d s_{\rm MF}(\varepsilon,m_z)/\d\varepsilon$.

The condition of Eq.~(\ref{eq:s_instability}) is not altered;
if the matrix $${\cal L}(\beta_{\rm MF}(\varepsilon,m_z)U_{\rm max},\vec{m}_{\rm eq}(\varepsilon,m_z))$$ has a negative eigenvalue, 
it is concluded that $s(\varepsilon,m_z)<s_{\rm MF}(\varepsilon,m_z)$.

When $\lambda_{\perp}>\lambda_z$, the mean-field entropy $s_{\rm MF}(\varepsilon,m_z)$ 
is a concave function of $\varepsilon$ and $m_z$, see Eq.~(\ref{eq:s_XXZ}).
Therefore, because of the inequality~(\ref{eq:s_inequality2-2}), $s_{\rm MF}(\varepsilon,m_z)=s_{\rm MF}^{**}(\varepsilon,m_z)$ in this case.
Since the canonical free energy is obtained by the microcanonical entropy, it is also concluded that $f(\beta,m_z)=f_{\rm MF}(\beta,m_z)$
when $\lambda_{\perp}>\lambda_z$.
The exactness of the mean-field theory holds 
when $\lambda_{\perp}>\lambda_z$ both in the canonical and in the microcanonical ensemble with a fixed value of the magnetization $m_z$.

\begin{figure*}[tbh]
\begin{center}
\begin{tabular}{ccc}
(a)& &(b)\\
\includegraphics[clip,width=4cm]{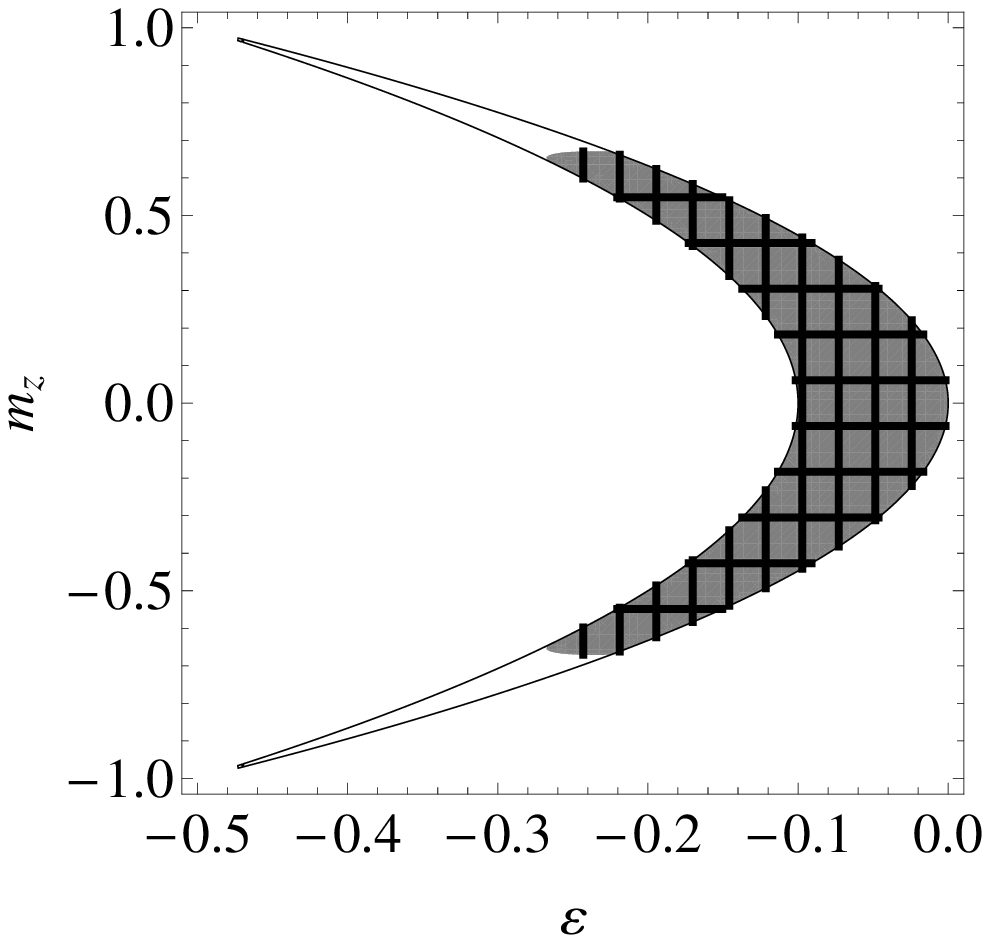}& &
\includegraphics[clip,width=4cm]{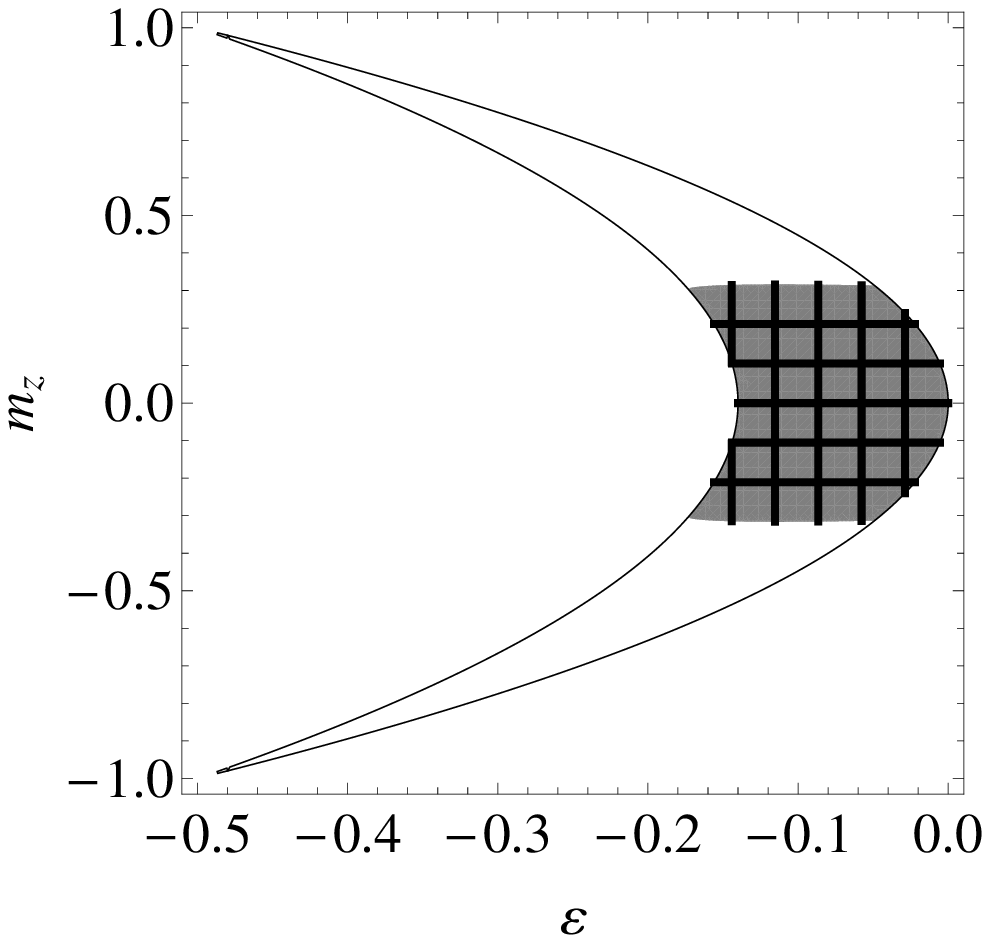} \\
(c)& &(d)\\
\includegraphics[clip,width=4cm]{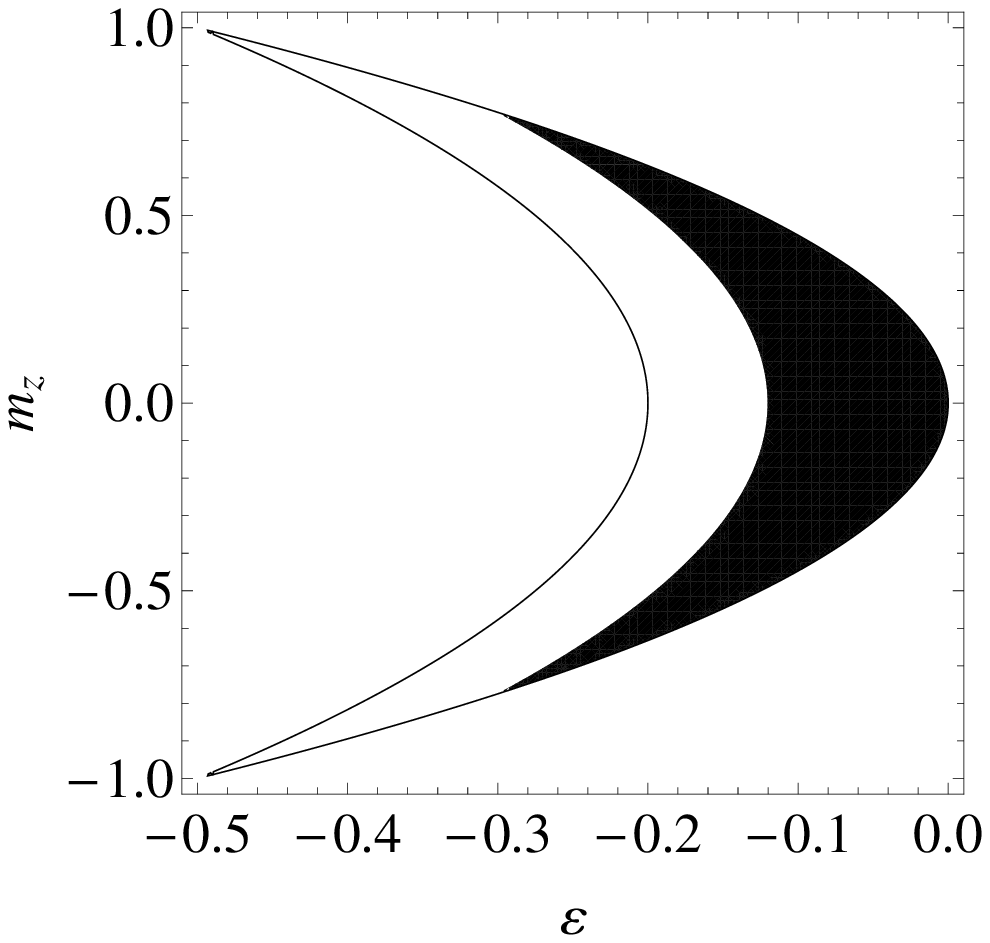}& &
\includegraphics[clip,width=4cm]{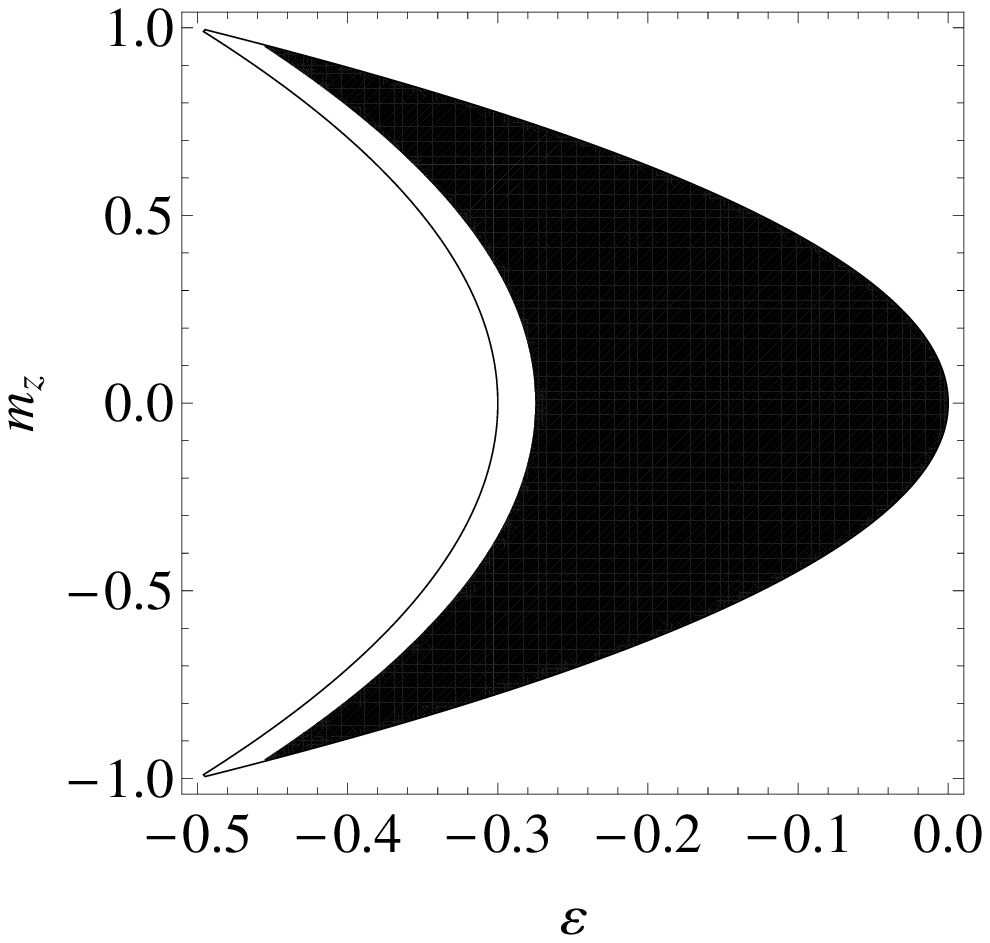}
\end{tabular}
\caption{The abscissa is the energy per spin $\varepsilon$ and the ordinate is $m_z$.
The region of ${\cal D}_{\rm MF}$ is inside the black solid line.
In the black region, $\beta_{\rm MF}(\varepsilon,m_z)<\tilde{\beta}/U_{\rm max}=1/(U_{\rm max}\lambda_z)$ and the mean-field theory is exact.
In the meshed gray region, ${\cal L}(\beta_{\rm MF}(\varepsilon,m_z)U_{\rm max},\vec{m}_{\rm eq}(\varepsilon,m_z))$ has a negative eigenvalue,
and the exactness of the mean-field theory is violated.
We cannot determine whether the mean-field theory is exact in the white region inside the solid line.
The parameters $\lambda_z$ and $U_{\rm max}$ are set to be 1 and 0.3, respectively.
(a) $\lambda_{\perp}=0.2$, (b) $\lambda_{\perp}=0.28$, (c) $\lambda_{\perp}=0.4$, and (d) $\lambda_{\perp}=0.6$.}
\label{fig:XXZ_microcanonical}
\end{center}
\end{figure*}

On the other hand, the result is non-trivial when $\lambda_z>\lambda_{\perp}$.
In Fig.~\ref{fig:XXZ_microcanonical}, 
we plot the region of ${\cal D}_{\rm MF}$ (inside the black solid line), 
$\beta_{\rm MF}(\varepsilon,m_z)<\tilde{\beta}/U_{\rm max}=1/(U_{\rm max}\lambda_z)$ (black region),
and the region where $${\cal L}(\beta_{\rm MF}(\varepsilon,m_z)U_{\rm max},\vec{m}_{\rm eq}(\varepsilon,m_z))$$ 
has a negative eigenvalue (meshed gray region),
in the $(\varepsilon,m_z)$ plane.
In the black region, the mean-field theory is exact.
In the meshed gray region, the mean-field theory is not exact.
In the white region inside the black solid line, we cannot determine whether the mean-field theory is exact from our analysis,
although the equilibrium states predicted by the mean-field model is at least locally stable.

\section{Summary and discussion}
\label{sec:summary}

In this paper, we analyzed quantum spin systems with long-range interactions.
We reduced the calculation of the free energy to the minimization of the free energy functional, which is expressed only by the classical variables.
We obtained some results from this variational expression of the free energy.
One of them is that the free energy is always identical to that of the mean-field model when there is no conserved quantity.
When there is a conserved quantity such as $m_z$, the mean-field model is exact at least for $\beta U_{\rm max}\leq\tilde{\beta}$.
The key quantity $U_{\rm max}$ is determined by the form of the interaction potential $\phi(\gamma\bm{r})$ and the spatial dimension $d$.
Moreover, if the matrix ${\cal L}(\beta U_{\rm max},\vec{m}_{\rm eq}(\beta,m_z))$ has a negative eigenvalue,
the state with the uniform magnetization profile predicted by the mean-field model is locally unstable,
and inhomogeneity which is not predicted by the mean-field model arises.
These results are the same as those in classical spin systems with long-range interactions.

We analyzed the systems also in the microcanonical ensemble, and obtained the same results as those in classical spin systems.
When the magnetization is not conserved,
the microcanonical entropy is identical to that of the mean-field model 
as long as the canonical and the microcanonical ensembles are equivalent in the mean-field model.
However, if these two ensembles are inequivalent in the mean-field model,
the microcanonical entropy may depend on the interaction potential $\phi(\bm{x})$.
The obtained results in this case are as follows. 
The mean-field model is exact if $\beta_{\rm MF}(\varepsilon)U_{\rm max}<\tilde{\beta}$
for all $\varepsilon$ such that $s_{\rm MF}(\varepsilon)<s_{\rm MF}^{**}(\varepsilon)$.
On the other hand, when ${\cal L}(\beta_{\rm MF}(\varepsilon)U_{\rm max},\vec{m}_{\rm eq}(\varepsilon))$ has a negative eigenvalue,
the microcanonical entropy is not identical to that of the mean-field model because the uniform state predicted by the mean-field model is locally unstable.
The analogous result holds when the magnetization $m_z$ is conserved.

As a concrete example, we investigated the spin-1/2 XXZ model with a long-range interaction.
When the magnetization is not conserved, the result is trivial (the mean-field theory is exact) both for the canonical and the microcanonical ensembles.
Therefore, we studied the model with a conserved $m_z$.
As a result, it is shown that the exactness of the mean-field theory is violated in a certain parameter region for both the canonical and the microcanonical ensembles.

In cold atom experiments, we can set up the system which is well isolated from the environment, 
but the $z$-component of the magnetization is not a conserved quantity in general.
Such a system will be described by the microcanonical ensemble without restriction of the value of $m^z$.
In this ensemble, the spin-1/2 XXZ model with long-range interactions is equivalent to the corresponding mean-field model, as mentioned above.
If we seek to observe the violation of exactness of the mean-field theory in long-range interacting systems,
it is necessary to consider another model.

In order to observe the violation of the exactness of the mean-field theory in the microcanonical ensemble,
the mean-field model corresponding to the system must satisfy $s_{\rm MF}(\varepsilon)\neq s_{\rm MF}^{**}(\varepsilon)$.
This condition means that the canonical and the microcanonical ensembles are inequivalent in the mean-field model
\footnote{Similarly, in order to observe the violation of the exactness of the mean-field theory in the canonical ensemble with a restriction of the magnetization value,
the mean-field model must satisfy $f_{\rm MF}(\beta,m_z)\neq f_{\rm MF}^{**}(\beta,m_z)$.
It implies that the canonical ensemble with a fixed value of the magnetization is not equivalent to the canonical ensemble with a fixed value of the magnetic field $h_z$.}.
Therefore, if the actual interaction is not infinite-range (or Curie-Weiss type),
not only the ensemble inequivalence predicted by analysis of the mean-field model,
but also the violation of the exactness of the mean-field theory will be observed.
Although it is known that the mean-field models reproduce many properties of general long-range interacting systems even quantitatively,
the infinite-range (or Curie-Weiss type) interaction does not become an adequate idealization of actual long-range interactions in such a situation.

Kastner pointed out that the inequivalence of the canonical and the microcanonical ensembles can be observed in a long-range interacting system
undergoing a temperature-driven first order phase transition~\cite{Kastner2010}.
A typical example of such spin models which possibly exhibits the violation of the exactness of the mean-field theory is the spin-1 Ising model with anisotropy,
\beq
H=-\frac{1}{2}\sum_{ij}\gamma^d\phi(\gamma\bm{r}_{ij})s_i^zs_j^z+D\sum_i(s_i^z)^2.
\label{eq:D_Ising}
\eeq
The corresponding mean-field model, which is obtained by replacing $\gamma^d\phi(\gamma\bm{r}_{ij})$ by $1/N$,
undergoes the temperature-driven first order phase transition for certain range of the parameter $D$.
However, this model is a special one in the sense that it is a classical spin system.
It is expected that more general quantum spin systems with the anisotropic term,
\beq
H=-\frac{1}{2s^2}\sum_{ij}\gamma^d\phi(\gamma \bm{r}_{ij})\sum_{a=x,y,z}\lambda_as_i^as_j^a+D\sum_i(s_i^z)^2.
\eeq
behaves similarly,
although it is not proven straightforwardly from the formalism given in this paper.
The presence of the nonlinear external field $D\sum_i(s_i^z)^2$ makes it difficult to treat by the formalism given in this paper.
A future work should treat nonlinear external fields such as $D\sum_i(s_i^z)^2$,
which will promote an experimental realization of characteristic properties of long-range interacting systems.

\acknowledgements
The author thanks Seiji Miyashita for valuable discussion and acknowledges JSPS for financial support (Grant No. 227835).

\appendix
\section{Calculation of $W(J_p)$}
\label{sec:W}

We seek to calculate the weight function $W(J_p)$.
From the definition, $W(J_p)$ is the total number of trajectories with length $l^d$ from 0 to $J_p$.
We can obtain the expression of $W(J_p)$ by counting such trajectories,
but we can do it more easily.
After all, $W(J_p)$ is the number of states with the total spin $J_p$ and the total magnetization $M_p=J_p$.
The states with $M_p=X$ have $J_p=X$ or $J_p\geq X+1$.
States with $J_p\geq X+1$ and $M_p=X$ are obtained by multiplying $S^-$ to states with $M_p=X+1$.
Therefore, if we refer to $W_M(X)$ as the number of states with $M_p=X$, the weight function is given by
\beq
W(X)=W_M(X)-W_M(X+1).
\eeq

For instance, in the case of spin-1/2, $W_M(X)$ is given by
\beq
W_M(X)=\frac{l^d!}{\left(\frac{l^d}{2}+X\right)!\left(\frac{l^d}{2}-X\right)!}.
\eeq
Therefore,
\begin{align}
W(J_p)&=W_M(J_p)-W_M(J_p+1)
\nonumber \\
&=\frac{2J_p+1}{\frac{l^d}{2}+J_p+1}\frac{l^d!}{\left(\frac{l^d}{2}+J_p\right)!\left(\frac{l^d}{2}-J_p\right)!}.
\end{align}
Applying the Stirling formula, we obtain
\begin{align}
\sigma(\vec{S})&={\rm Lim}\frac{1}{l^d}\ln W(J_p)
\nonumber \\
&=-\frac{1+|\vec{S}|}{2}\ln\frac{1+|\vec{S}|}{2}-\frac{1-|\vec{S}|}{2}\ln\frac{1-|\vec{S}|}{2}.
\label{eq:entropy_spin1/2}
\end{align}

The weight function and the entropy function are calculated similarly for higher spins ($s=1,3/2,\dots$),
though the calculation becomes complex.
In the case of $s=1$, the result is
\beq
W_M(X)=\sum_{N=0}^{l^d-X}\frac{l^d}{[(l^d-N+X)/2]![(l^d-N-X)/2]!N!},
\eeq
and
\begin{align}
\sigma(\vec{S})=&-\frac{1-x+|\vec{S}|}{2}\ln\frac{1-x+|\vec{S}|}{2}
\nonumber \\
&-\frac{1-x-|\vec{S}|}{2}\ln\frac{1-x-|\vec{S}|}{2}-x\ln x,
\end{align}
where $$x=\frac{-1+\sqrt{4-3|\vec{S}|^2}}{3}.$$

\section{Proof of Eq.~(\ref{eq:Delta})}
\label{sec:Delta}

We derive the inequality~(\ref{eq:Delta}).
For simplification, we write $\vec{e}\in\vec{\omega}_p(\Delta)$ if we can express the unit vector $\vec{e}$ by
$\vec{e}=(\sin\theta\cos\phi,\sin\theta\sin\phi,\cos\theta)$ with $(\theta,\phi)\in\omega_p(\Delta)$.

First, we show that $|\vec{a}-\vec{b}|\leq 2\Delta$ for arbitrary unit vectors $\vec{a}, \vec{b}\in \vec{\omega}_p(\Delta)$.
We have
\beq
|\vec{a}-\vec{b}|=|(\vec{a}-\vec{e^*}_p)-(\vec{b}-\vec{e^*}_p)|\leq |\vec{a}-\vec{e^*}_p|+|\vec{b}-\vec{e^*}_p|.
\label{eq:ab}
\eeq
We choose the $z$ axis as the direction of the vector $\vec{e^*}_p$.
Any $\vec{x}\in\vec{\omega}_p(\Delta)$ is represented as
$\vec{x}=(\sin\theta\cos\phi,\sin\theta\sin\phi,\cos\theta)$ with $0\leq\theta\leq\Delta$,
which yields $|\vec{x}-\vec{e^*}_p|=2\sin(\theta/2)\leq\theta\leq \Delta$.
Therefore, from Eq.~(\ref{eq:ab}), we have
\beq
|\vec{a}-\vec{b}|\leq |\vec{a}-\vec{e^*}_p|+|\vec{b}-\vec{e^*}_p|\leq 2\Delta.
\eeq

We now show Eq.~(\ref{eq:Delta}).
From the definition, we have
\begin{align}
&\sup_{(\theta_p,\phi_p)\in\omega_p(\Delta)}\tilde{H}^{\rm cl}(\{ J_p,\theta_p,\phi_p\})-E^*(\{ J_p\})
\nonumber \\
&=\sup_{(\theta_p,\phi_p)\in\omega_p(\Delta)}\left[-\frac{1}{2s^2}\sum_{p,q}^{(L/l)^d}\gamma^d\phi_{pq}J_pJ_q\right.
\nonumber \\
&\left.\qquad\times\sum_{a=x,y,z}\lambda_a((\vec{e}_p)^a(\vec{e}_q)^a-(\vec{e^*}_p)^a(\vec{e^*}_q)^a)\right]
\nonumber \\
&\leq\sup_{(\theta_p,\phi_p)\in\omega_p(\Delta)}\frac{1}{2s^2}\sum_{p,q}^{(L/l)^d}\gamma^d\phi_{pq}J_pJ_q\lambda_{\rm max}
\nonumber \\
&\qquad\times|\vec{e}_p\cdot\vec{e}_q-\vec{e^*}_p\cdot\vec{e^*}_q|.
\end{align}
Here we define $\vec{e}_p=(\sin\theta_p\cos\phi_p,\sin\theta_p\sin\phi_p,\cos\theta_p)$ and
$\vec{e^*}_p=(\sin\theta_p^*\cos\phi_p^*,\sin\theta_p^*\sin\phi_p^*,\cos\phi_p^*)$.
Because $J_p\leq sl^d$ and
\begin{align}
|\vec{e}_p\cdot\vec{e}_q-\vec{e^*}_p\cdot\vec{e^*}_q|
&= |(\vec{e}_p-\vec{e^*}_p)\cdot\vec{e}_q+\vec{e^*}_p\cdot(\vec{e}_q-\vec{e^*}_q)|
\nonumber \\
&\leq |\vec{e}_p-\vec{e^*}_p|+|\vec{e}_q-\vec{e^*}_q|
\nonumber \\
&\leq 2\Delta,
\end{align}
we obtain the inequality~(\ref{eq:Delta}),
\begin{align}
\sup_{(\theta_p,\phi_p)\in\omega_p(\Delta)}\tilde{H}^{\rm cl}(\{ J_p,\theta_p,\phi_p\})-E^*(\{ J_p\})
\nonumber \\
\leq\lambda_{\rm max}\Delta l^{2d}\sum_{pq}\gamma^d\phi_{pq}
=\lambda_{\rm max}\Delta L^d.
\end{align}

\section{Proof of Eq.~(\ref{eq:tilde_XXZ})}
\label{sec:tilde_XXZ}

In the model described by Eq.~(\ref{eq:H_XXZ}), the mean-field free energy is given by Eq.~(\ref{eq:f_XXZ}).
The Hessian matrix ${\cal L}(\beta,\vec{m})$ defined by Eq.~(\ref{eq:L}) determines the convexity of the mean-field free energy.
If the minimum eigenvalue of ${\cal L}(\beta,\vec{m})$ is positive for all $\vec{m}$, 
the mean-field free energy is convex for $\vec{m}$ and $f_{\rm MF}(\beta,\vec{m})=f_{\rm MF}^{**}(\beta,\vec{m})$.
Therefore, $\tilde{\beta}$ corresponds to $\beta$ such that the minimum eigenvalue of ${\cal L}(\beta,\vec{m})$ becomes zero at some $\vec{m}$.

Eigenvalues of ${\cal L}(\beta,\vec{m})$ are calculated and given by
\begin{align}
&l_0(\beta,\vec{m})\equiv -\lambda_{\perp}+\frac{1}{2\beta}\frac{1}{|\vec{m}|}\ln\frac{1+|\vec{m}|}{1-|\vec{m}|}, \\
&l_{\pm}(\beta,\vec{m})\equiv \frac{1}{2\beta}\frac{1}{|\vec{m}|}\ln\frac{1+|\vec{m}|}{1-|\vec{m}|}
\nonumber \\
&+\frac{1}{2}\bigg[-\lambda_{\perp}-\lambda_z+B|\vec{m}|^2
\nonumber \\
&\pm\rt{(\lambda_{\perp}-\lambda_z+B|\vec{m}|^2)^2
+4(\lambda_z-\lambda_{\perp})B(|\vec{m}|^2-m_z^2)}\bigg],
\label{eq:l-}
\end{align}
where $$B\equiv\frac{1}{2\beta}\frac{1}{|\vec{m}|^2}\left[\frac{2}{1-|\vec{m}|^2}-\frac{1}{|\vec{m}|}\ln\frac{1+|\vec{m}|}{1-|\vec{m}|}\right].$$
By using these eigenvalues, $\tilde{\beta}$ is given by
\beq
\tilde{\beta}=\sup\left\{\beta>0:\min\left[\inf_{\vec{m}}l_0(\beta,\vec{m}),\inf_{\vec{m}}l_-(\beta,\vec{m})\right]>0\right\}.
\eeq

Obviously, we find
\beq
\inf_{\vec{m}}l_0(\beta,\vec{m})=-\lambda_{\perp}+\frac{1}{\beta}.
\eeq
When $\lambda_{\perp}>\lambda_z$,
it is shown that $l_-(\beta,\vec{m})\geq l_0(\beta,\vec{m})$.
Therefore, $\tilde{\beta}=1/\lambda_{\perp}$ in this case.

Next we consider the case of $\lambda_z\geq\lambda_{\perp}$.
We rewrite $l_-(\beta,\vec{m})$ as
\begin{align}
l_-(\beta,\vec{m})&=
\frac{1}{2\beta}\frac{1}{|\vec{m}|}\ln\frac{1+|\vec{m}|}{1-|\vec{m}|}
\nonumber \\
&+\frac{1}{2}\bigg[-\lambda_{\perp}-\lambda_z+B|\vec{m}|^2
\nonumber \\
&-\rt{(\lambda_{\perp}-\lambda_z-B|\vec{m}|^2)^2
-4(\lambda_z-\lambda_{\perp})Bm_z^2}\bigg].
\end{align}
From this expression,
\beq
l_-(\beta,\vec{m})\geq-\lambda_z+\frac{1}{2\beta}\frac{1}{|\vec{m}|}\ln\frac{1+|\vec{m}|}{1-|\vec{m}|}.
\eeq
Hence, we have
\beq
\inf_{\vec{m}}l_-(\beta,\vec{m})\geq -\lambda_z+\frac{1}{\beta}.
\label{eq:l-1}
\eeq
On the other hand, from Eq.~(\ref{eq:l-}),
\begin{align}
l_-(\beta,\vec{m})&\leq\frac{1}{2\beta}\frac{1}{|\vec{m}|}\ln\frac{1+|\vec{m}|}{1-|\vec{m}|}-\lambda_z+B|\vec{m}|^2
\nonumber \\
&=-\lambda_z+\frac{1}{\beta}\frac{1}{1-|\vec{m}|^2}.
\end{align}
Thus we have
\beq
\inf_{\vec{m}}l_-(\beta,\vec{m})\leq -\lambda_z+\frac{1}{\beta}.
\label{eq:l-2}
\eeq
From Eqs.~(\ref{eq:l-1}) and (\ref{eq:l-2}), we obtain
\beq
\inf_{\vec{m}}l_-(\beta,\vec{m})=-\lambda_z+\frac{1}{\beta}.
\eeq
Therefore, when $\lambda_z\geq\lambda_{\perp}$, we have $\tilde{\beta}=1/\lambda_z$.

The analysis in this appendix concludes that $\tilde{\beta}=\min(1/\lambda_{\perp},1/\lambda_z)$.


\begin{thebibliography}{20}
\expandafter\ifx\csname natexlab\endcsname\relax\def\natexlab#1{#1}\fi
\expandafter\ifx\csname bibnamefont\endcsname\relax
  \def\bibnamefont#1{#1}\fi
\expandafter\ifx\csname bibfnamefont\endcsname\relax
  \def\bibfnamefont#1{#1}\fi
\expandafter\ifx\csname citenamefont\endcsname\relax
  \def\citenamefont#1{#1}\fi
\expandafter\ifx\csname url\endcsname\relax
  \def\url#1{\texttt{#1}}\fi
\expandafter\ifx\csname urlprefix\endcsname\relax\def\urlprefix{URL }\fi
\providecommand{\bibinfo}[2]{#2}
\providecommand{\eprint}[2][]{\url{#2}}

\bibitem[{\citenamefont{Campa et~al.}(2009)\citenamefont{Campa, Dauxois, and
  Ruffo}}]{Campa_review}
\bibinfo{author}{\bibfnamefont{A.}~\bibnamefont{Campa}},
  \bibinfo{author}{\bibfnamefont{T.}~\bibnamefont{Dauxois}}, \bibnamefont{and}
  \bibinfo{author}{\bibfnamefont{S.}~\bibnamefont{Ruffo}},
  \bibinfo{journal}{Phys. Rep.} \textbf{\bibinfo{volume}{480}},
  \bibinfo{pages}{57} (\bibinfo{year}{2009}).

\bibitem[{\citenamefont{Dauxois et~al.}(2008)\citenamefont{Dauxois, Ruffo, and
  Cugliandolo}}]{Les_Houches_long}
\bibinfo{author}{\bibfnamefont{T.}~\bibnamefont{Dauxois}},
  \bibinfo{author}{\bibfnamefont{S.}~\bibnamefont{Ruffo}}, \bibnamefont{and}
  \bibinfo{author}{\bibfnamefont{L.}~\bibnamefont{Cugliandolo}}, in
  \emph{\bibinfo{booktitle}{Les Houches Summer School}} (\bibinfo{year}{2008}).


\bibitem[{\citenamefont{O'Dell et~al.}(2000)\citenamefont{O'Dell, Giovanazzi,
  Kurizki, and Akulin}}]{ODell2000}
\bibinfo{author}{\bibfnamefont{D.}~\bibnamefont{O'Dell}},
  \bibinfo{author}{\bibfnamefont{S.}~\bibnamefont{Giovanazzi}},
  \bibinfo{author}{\bibfnamefont{G.}~\bibnamefont{Kurizki}}, \bibnamefont{and}
  \bibinfo{author}{\bibfnamefont{V.~M.} \bibnamefont{Akulin}},
  \bibinfo{journal}{Phys. Rev. Lett.} \textbf{\bibinfo{volume}{84}},
  \bibinfo{pages}{5687} (\bibinfo{year}{2000}).

\bibitem[{\citenamefont{Kastner}(2010{\natexlab{a}})}]{Kastner2010}
\bibinfo{author}{\bibfnamefont{M.}~\bibnamefont{Kastner}},
  \bibinfo{journal}{Phys. Rev. Lett.} \textbf{\bibinfo{volume}{104}},
  \bibinfo{pages}{240403} (\bibinfo{year}{2010}{\natexlab{a}}).

\bibitem[{\citenamefont{Kastner}(2010{\natexlab{b}})}]{Kastner_statmech2010}
\bibinfo{author}{\bibfnamefont{M.}~\bibnamefont{Kastner}}, \bibinfo{journal}{J.
  Stat. Mech.} \textbf{\bibinfo{volume}{2010}}, \bibinfo{pages}{P07006}
  (\bibinfo{year}{2010}{\natexlab{b}}).

\bibitem[{\citenamefont{Cannas et~al.}(2000)\citenamefont{Cannas, de~Magalhaes,
  and Tamarit}}]{Cannas2000}
\bibinfo{author}{\bibfnamefont{S.~A.} \bibnamefont{Cannas}},
  \bibinfo{author}{\bibfnamefont{A.~C.~N.} \bibnamefont{de~Magalhaes}},
  \bibnamefont{and} \bibinfo{author}{\bibfnamefont{F.~A.}
  \bibnamefont{Tamarit}}, \bibinfo{journal}{Phys. Rev. B}
  \textbf{\bibinfo{volume}{61}}, \bibinfo{pages}{11521} (\bibinfo{year}{2000}).

\bibitem[{\citenamefont{Tamarit and Anteneodo}(2000)}]{Tamarit2000}
\bibinfo{author}{\bibfnamefont{F.}~\bibnamefont{Tamarit}} \bibnamefont{and}
  \bibinfo{author}{\bibfnamefont{C.}~\bibnamefont{Anteneodo}},
  \bibinfo{journal}{Phys. Rev. Lett.} \textbf{\bibinfo{volume}{84}},
  \bibinfo{pages}{208} (\bibinfo{year}{2000}).

\bibitem[{\citenamefont{Campa et~al.}(2003)\citenamefont{Campa, Giansanti, and
  Moroni}}]{Campa2003}
\bibinfo{author}{\bibfnamefont{A.}~\bibnamefont{Campa}},
  \bibinfo{author}{\bibfnamefont{A.}~\bibnamefont{Giansanti}},
  \bibnamefont{and} \bibinfo{author}{\bibfnamefont{D.}~\bibnamefont{Moroni}},
  \bibinfo{journal}{J. Phys A: Math. Gen.} \textbf{\bibinfo{volume}{36}},
  \bibinfo{pages}{6897} (\bibinfo{year}{2003}).

\bibitem[{\citenamefont{Barr{\'e} et~al.}(2005)\citenamefont{Barr{\'e},
  Bouchet, Dauxois, and Ruffo}}]{Barre2005}
\bibinfo{author}{\bibfnamefont{J.}~\bibnamefont{Barr{\'e}}},
  \bibinfo{author}{\bibfnamefont{F.}~\bibnamefont{Bouchet}},
  \bibinfo{author}{\bibfnamefont{T.}~\bibnamefont{Dauxois}}, \bibnamefont{and}
  \bibinfo{author}{\bibfnamefont{S.}~\bibnamefont{Ruffo}}, \bibinfo{journal}{J.
  Stat. Phys.} \textbf{\bibinfo{volume}{119}}, \bibinfo{pages}{677}
  (\bibinfo{year}{2005}).

\bibitem[{\citenamefont{Mori}(2010)}]{Mori2010_analysis}
\bibinfo{author}{\bibfnamefont{T.}~\bibnamefont{Mori}}, \bibinfo{journal}{Phys.
  Rev. E} \textbf{\bibinfo{volume}{82}}, \bibinfo{pages}{060103}
  (\bibinfo{year}{2010}).

\bibitem[{\citenamefont{Mori}(2011)}]{Mori2011_instability}
\bibinfo{author}{\bibfnamefont{T.}~\bibnamefont{Mori}}, \bibinfo{journal}{Phys.
  Rev. E} \textbf{\bibinfo{volume}{84}}, \bibinfo{pages}{031128}
  (\bibinfo{year}{2011}{\natexlab{a}}).
  
\bibitem[{\citenamefont{Mori}(2012)}]{Mori2012_microcanonical}
\bibinfo{author}{\bibfnamefont{T.}~\bibnamefont{Mori}},
  \bibinfo{journal}{J. Stat. Phys.} \textbf{\bibinfo{volume}{147}},
  \bibinfo{pages}{1020} (\bibinfo{year}{2012}).

\bibitem[{\citenamefont{Lebowitz and Penrose}(1966)}]{Lebowitz-Penrose1966}
\bibinfo{author}{\bibfnamefont{J.}~\bibnamefont{Lebowitz}} \bibnamefont{and}
  \bibinfo{author}{\bibfnamefont{O.}~\bibnamefont{Penrose}},
  \bibinfo{journal}{J. Math. Phys.} \textbf{\bibinfo{volume}{7}},
  \bibinfo{pages}{98} (\bibinfo{year}{1966}).
  
\bibitem[{\citenamefont{Lieb}(1966)}]{Lieb1966}
\bibinfo{author}{\bibfnamefont{E.~H.}~\bibnamefont{Lieb}},
  \bibinfo{journal}{J. Math. Phys.} \textbf{\bibinfo{volume}{7}},
  \bibinfo{pages}{1016} (\bibinfo{year}{1966}).
  
\bibitem[{\citenamefont{Brout}(1960)}]{Brout1960}
\bibinfo{author}{\bibfnamefont{R.}~\bibnamefont{Brout}},
  \bibinfo{journal}{Phys. Rev.} \textbf{\bibinfo{volume}{118}},
  \bibinfo{pages}{1009} (\bibinfo{year}{1960}).

\bibitem[{\citenamefont{Bricmont and Fontaine}(1982)}]{Bricmont1982}
\bibinfo{author}{\bibfnamefont{J.}~\bibnamefont{Bricmont}} \bibnamefont{and}
  \bibinfo{author}{\bibfnamefont{J.~R.}~\bibnamefont{Fontaine}},
  \bibinfo{journal}{J. Comm. Math. Phys.} \textbf{\bibinfo{volume}{86}},
  \bibinfo{pages}{337} (\bibinfo{year}{1982}).

\bibitem[{\citenamefont{Golden}(1965)}]{Golden1965}
\bibinfo{author}{\bibfnamefont{S.}~\bibnamefont{Golden}},
  \bibinfo{journal}{Phys. Rev.} \textbf{\bibinfo{volume}{137}},
  \bibinfo{pages}{B1127} (\bibinfo{year}{1965}).

\bibitem[{\citenamefont{Thompson}(1965)}]{Thompson1965}
\bibinfo{author}{\bibfnamefont{C.}~\bibnamefont{Thompson}},
  \bibinfo{journal}{J. Math. Phys.} \textbf{\bibinfo{volume}{6}},
  \bibinfo{pages}{1812} (\bibinfo{year}{1965}).

\bibitem[{\citenamefont{Lieb}(1973)}]{Lieb1973}
\bibinfo{author}{\bibfnamefont{E.~H.} \bibnamefont{Lieb}},
  \bibinfo{journal}{Comm. Math. Phys.} \textbf{\bibinfo{volume}{31}},
  \bibinfo{pages}{327} (\bibinfo{year}{1973}).
  
\bibitem[{\citenamefont{Gates and Penrose}(1969)}]{Gates1969}
\bibinfo{author}{\bibfnamefont{D.}~\bibnamefont{Gates}} \bibnamefont{and}
  \bibinfo{author}{\bibfnamefont{O.}~\bibnamefont{Penrose}},
  \bibinfo{journal}{J. Comm. Math. Phys.} \textbf{\bibinfo{volume}{15}},
  \bibinfo{pages}{255} (\bibinfo{year}{1969}).
  
\bibitem[{\citenamefont{Gates and Penrose}(1970)}]{Gates1970}
\bibinfo{author}{\bibfnamefont{D.}~\bibnamefont{Gates}} \bibnamefont{and}
  \bibinfo{author}{\bibfnamefont{O.}~\bibnamefont{Penrose}},
  \bibinfo{journal}{J. Comm. Math. Phys.} \textbf{\bibinfo{volume}{16}},
  \bibinfo{pages}{231} (\bibinfo{year}{1970}).
  
\bibitem[{\citenamefont{Kiessling and Percus}(1995)}]{Kiessling1995}
\bibinfo{author}{\bibfnamefont{M.~K.~H.}~\bibnamefont{Kiessling}} \bibnamefont{and}
  \bibinfo{author}{\bibfnamefont{J.~K.}~\bibnamefont{Percus}},
  \bibinfo{journal}{J. Stat. Phys.} \textbf{\bibinfo{volume}{78}},
  \bibinfo{pages}{1337} (\bibinfo{year}{1995}).


\end{thebibliography}

\end{document}